%% file: main.tex
\documentclass[pdflatex,sn-mathphys-num]{sn-jnl}%

\makeatletter
\@twosidefalse
\@mparswitchfalse
\makeatother
\usepackage{graphicx}%
\usepackage{multirow}%
\usepackage{amsmath,amssymb,amsfonts}%
\usepackage{amsthm}%
\usepackage{mathrsfs}%
\usepackage[title]{appendix}%
\usepackage{xcolor}%
\usepackage{textcomp}%
\usepackage{manyfoot}%
\usepackage{booktabs}%
\usepackage{algorithm}%
\usepackage{algorithmicx}%
\usepackage{algpseudocode}%
\usepackage{listings}%
\usepackage{bibunits}
\usepackage{longtable}
\usepackage{capt-of}
\usepackage{placeins}

\theoremstyle{thmstyleone}%

\theoremstyle{thmstyletwo}%

\theoremstyle{thmstylethree}%

\begin{document}

\title[Article Title]{Advancing global crop simulations with differentiable modelling} %

\author*[1,2]{\fnm{Yunan} \sur{Lin}}\email{yunan.lin@tum.de}

\author[1,2]{\fnm{Sebastian} \sur{Bathiany}}

\author[2,1]{\fnm{Maximilian} \sur{Gelbrecht}}

\author[2]{\fnm{Jens} \sur{Heinke}}

\author[2]{\fnm{Christoph} \sur{Müller}}

\author*[1,2,3]{\fnm{Niklas} \sur{Boers}}\email{n.boers@tum.de}

\affil[1]{\orgname{Munich Climate Center and Earth System Modelling Group, Department of Aerospace and Geodesy, School of Engineering and Design, Technical University of Munich}, \city{Munich}, \country{Germany}}

\affil[2]{\orgname{Potsdam Institute for Climate Impact Research, Member of the Leibniz Association}, \city{Potsdam}, \country{Germany}}

\affil[3]{\orgname{University of Exeter}, \city{Exeter}, \country{UK}}

\abstract{

Global gridded crop models (GGCMs) are crucial for assessing the performance of food producing systems, for example under climate change or their contribution to environmental challenges. However, systematic calibration of their parameters remains a long-standing challenge. Persisting uncertainties in these parameters propagate into future crop projections, thereby affecting and potentially biasing food security assessments. Here we present a fully process-based differentiable GGCM implementation that offers a fundamentally new approach to parameter calibration, uncertainty quantification, yield loss attribution, and management optimization. Using automatic differentiation, we optimize selected model parameters against site-scale carbon and water flux observations, reducing normalized root-mean-square errors at held-out sites by 24.8--36.2\% compared to the default state-of-the-art model. Without calibration against yield data, the optimized model better captures detrended yield anomalies across 61.2--64.3\% of global crop area and reduces low-yield errors by 3.4--43.5\% for wheat, rice, maize and soybean. It also ranks among the best performers in comparison with GGCMI Phase 3 models. Future simulations show that calibration reduces the projected end-of-century rice production increase from 55.1\% to 22.1\% under SSP5--8.5, suggesting substantial overestimation by the default parameters. Parameter ensembles further reveal that parameterizations with comparable historical skill can produce markedly different future production responses. 
Beyond calibration, differentiability enables the attribution of crop losses under extreme events and supports the design of adaptive crop management strategies. For the 2018 European drought, we identify spring heat and water limitation as major contributors to wheat yield losses and develop more effective irrigation strategies for mitigating simulated regional yield losses. %

}

\maketitle

\section*{Introduction}

Climate change increasingly threatens global food production, with impacts varying among crops and regions \cite{jagermeyr2021climate, hultgren2025impacts, heikonen2025climate, yang2025climate}. Global gridded crop models (GGCMs) are central to assessing these risks, as they simulate the effects of climate and management on crop growth and yield \cite{jones2003dssat, bondeau2007modelling, von2018implementing, enders2023simplace, mialyk2024water}. However, projected climate responses differ widely among models owing to uncertainty in process representation and parameterization~\cite{muller2017evaluation, folberth2019parameterization, franke2020ggcmi, muller2021exploring, wang2024pathways, li2025global}. GGCMs also often underestimate low-yield risks and the magnitude of losses during climatic extremes \cite{kornhuber2023risks, auer2026global}. Together, these limitations reduce confidence in assessments of crop production risks and adaptation benefits under climate change. Improving the reliability of crop projections therefore requires progress in calibrating process-based crop models, quantifying parameter uncertainty and better diagnosing yield losses from climate extremes.

Process-based GGCMs link climate to crop growth and yield through coupled physiological and soil processes. The magnitude and timing of simulated responses to temperature, water, radiation and nutrient availability depend on many process parameters \cite{folberth2019parameterization}. Poorly constrained parameter values can bias simulated yield responses, making calibration against observations essential for reliable projections. Yet systematic calibration remains a long-standing challenge for GGCMs \cite{maiorano2017crop, xiong2016calibration, wallach2025there}. Current approaches rely on expert judgement, iterative trial and error and repeated forward searches, which become increasingly impractical as the number of parameters grows \cite{seidel2018towards, gawinowski2025selecting, machado2026large}. Moreover, these efforts typically target end-of-season yield rather than sub-seasonal processes, thereby leaving the underlying crop dynamics weakly constrained despite the increasing availability of growing-season carbon and water flux observations \cite{he2017data, seidel2018towards, pastorello2020fluxnet2015}.

Quantifying parameter uncertainty poses another major challenge \cite{wang2024pathways}. Multiple parameterizations perform comparably against historical observations but may diverge under future climate \cite{xiong2016calibration,he2017data}. Global crop projections from multi-model ensembles often use a single parameter set for each model, although parameter uncertainty within individual models can be substantial \cite{folberth2019parameterization, wallach2025there}. Constructing and evaluating parameter ensembles at the global scale is difficult because each ensemble member requires separate calibration. Given the projected increases in frequency and intensity of extreme events in a changing climate \cite{ipcc2023synthesis}, identifying the extreme weather drivers of yield losses is crucial for understanding crop vulnerability and informing adaptation. Existing GGCM sensitivity analyses rely on large ensembles of forward simulations, an approach that is both computationally intensive and restricted to predefined weather variables and perturbation ranges \cite{franke2020ggcmi}. GGCM emulators reduce computational cost but are less reliable outside their training ranges. Emulators trained on growing-season means do not resolve the sub-seasonal weather conditions or within-season timing of extreme events \cite{franke2020emulators}.

Differentiable modelling offers a promising approach to address these challenges directly within process-based models using gradient-based optimization~\cite{irrgang2021towards, gelbrecht2023differentiable, shen2023differentiable}, and has recently been applied in land surface and atmospheric modelling \cite{fang2026differentiable, kochkov2024neural}. Its core tool, automatic differentiation (AD), computes the local sensitivities of model outputs or optimization objectives to parameters and inputs~\cite{baydin2018automatic}. In reverse mode, a single backward pass yields gradients of the objective with respect to all selected parameters or inputs. Finite-difference methods instead require repeated forward simulations with slightly perturbed parameter values or inputs. Conventional GGCM implementations are not designed to propagate gradients through their complete seasonal simulations. An early proof-of-concept study applied AD to crop model sensitivity analysis~\cite{lauvernet2012using}. A recent hybrid GGCM used this approach to train embedded neural network components rather than optimize physical process parameters~\cite{lin2025neuralcrop}. To our knowledge, this gradient-based parameter optimization has not previously been demonstrated in a fully process-based GGCM. Here we present a differentiable implementation of the crop and soil formulations of LPJmL~\cite{bondeau2007modelling, 2018LPJmL, von2018implementing, lutz2019simulating, wirth2024biological} (Agrocosm), capable of quantifying the sensitivity of crop growth, soil dynamics and end-of-season yield to process parameters and daily weather inputs.

We use Agrocosm to demonstrate the advances enabled by differentiability in a fully process-based GGCM. We integrate process observations into GGCM calibration using AD to optimize model parameters against site-scale carbon and water fluxes. The resulting parameterizations generalize to held-out sites and strongly improve global simulations of historical yield variability for wheat, rice, maize and soybean, despite being calibrated without yield data. Future simulations under two emission scenarios show how parameter optimization alters projected production responses and yield sensitivities to CO$_2$ and temperature, while revealing substantial projection uncertainty across crops and regions. For rice under SSP5--8.5, parameter uncertainty accounts for about twice as much end-of-century projection variance as climate model uncertainty. We further use derivatives of end-of-season yield with respect to daily weather inputs to identify spring heat and water limitation as contributors to the 2018 European wheat yield loss. We also demonstrate how AD enables more effective irrigation strategies to improve regional yield recovery for the same event. Together, these analyses highlight how differentiability expands the capabilities of fully process-based GGCMs for assessing climate risks to crop production. 

\section*{Results}
\subsection*{Calibration using automatic differentiation} %

\begin{figure}[p]
\centering
\makebox[\textwidth][c]{%
\includegraphics[width=1.01\textwidth]{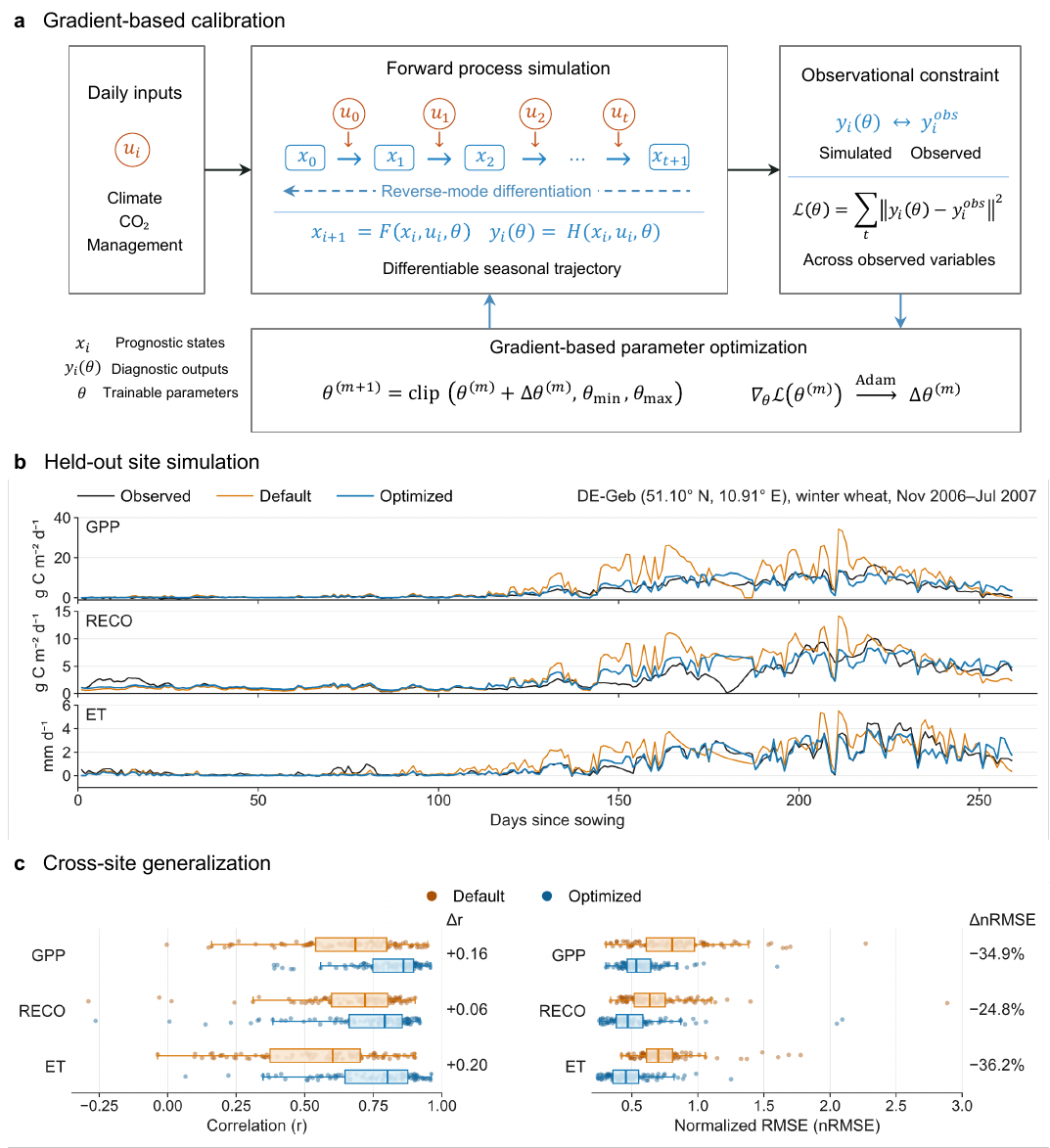}%
}
\caption{\textbf{Differentiable parameter calibration.}
\textbf{a}, Differentiable seasonal simulation, observational constraint and gradient-based parameter optimization. Reverse mode automatic differentiation computes gradients of the loss $\mathcal{L}$ with respect to the trainable parameters $\theta$. The Adam optimizer~\cite{kingma2015adam} uses these gradients to determine the direction and size of the parameter changes $\Delta\theta^{(m)}$ at calibration iteration $m$. Updated parameters are clipped to their prescribed bounds. $\mathbf{x}_i$, $\mathbf{u}_i$ and $\mathbf{y}_i$ denote prognostic states, daily external inputs and simulated diagnostic fluxes, respectively. $H$ maps model states and inputs to simulated diagnostic fluxes.
\textbf{b}, Daily gross primary production (GPP), ecosystem respiration (RECO) and evapotranspiration (ET) for winter wheat in central Germany (DE-Geb) in 2006. The trajectory provides an illustrative example to show how the optimized parameterization more closely reproduces the seasonal dynamics of all three fluxes at an unseen site.
\textbf{c}, Correlations and normalized root-mean-square error (nRMSE) between simulated fluxes and held-out observations of daily GPP, RECO and ET, evaluated across 100 growing seasons from 27 physical sites. Points denote individual growing seasons. Boxes show the median and interquartile range, with whiskers extending to 1.5 times the interquartile range. Annotations report changes after first averaging seasons within each site for each crop and then averaging the 27 physical sites. $\Delta r$ denotes the change in mean correlation across the 27 sites after parameter optimization.
}
\label{fig:differentiable_calibration}
\end{figure}

Agrocosm provides a differentiable and GPU-optimized implementation of the crop and soil formulations of LPJmL ~\cite{bondeau2007modelling, 2018LPJmL, von2018implementing, lutz2019simulating, wirth2024biological} for complete growing-season simulations in Julia. The reference simulations use the same implementation with default parameters adapted from LPJmL. Exploiting our implementation's differentiability, we compute reverse-mode gradients of the observation loss to optimize selected parameters against daily observations of gross primary production (GPP), ecosystem respiration (RECO) and evapotranspiration (ET) (\hyperref[sec:gradient-based_model_optimization]{Methods}; Fig.~\ref{fig:differentiable_calibration}a). Cross-site generalization is evaluated across 100 growing seasons from 27 flux-tower sites spanning wheat, maize, rice and soybean using grouped-site fivefold cross-validation, with entire physical sites held out (Supplementary Fig.~\ref{fig:flux_site_distribution}) \cite{pastorello2020fluxnet2015}. Sensitivity analyses identify a subset of crop and soil parameters for calibration against daily GPP, RECO and ET (\hyperref[sec:gradient-based_model_optimization]{Methods} and Supplementary Information section~\ref{sec:parameter_selection}). 

After calibration against flux-tower observations, mean temporal correlation across the 27 held-out sites increases from 0.65 to 0.81 for GPP, 0.65 to 0.71 for RECO, and 0.54 to 0.74 for ET, while normalized root-mean-square error (nRMSE) decreases by 34.9\%, 24.8\% and 36.2\%, respectively (Fig.~\ref{fig:differentiable_calibration}b). Correlation improves at 25, 24 and 27 of the 27 sites, and nRMSE improves at 25, 23 and 27 sites, for GPP, RECO and ET, respectively (Supplementary Table~\ref{tab:cv5_primary_results}). The improvements extend across all four crops. The crop-parameter stage accounts for most of the GPP and RECO gains, whereas the soil-parameter stage further improves ET correlation at every site (Supplementary Table~\ref{tab:cv5_stage_contributions}). This evaluation therefore supports transferable constraints on coupled carbon and water dynamics rather than improvements restricted to calibration sites. Using the held-out DE-Geb wheat season as an example (Fig.~\ref{fig:differentiable_calibration}c), the default model captures the broad seasonality of all three fluxes but exaggerates their growing-season magnitudes. Optimization substantially reduces these amplitude biases and more closely tracks observed within-season variations across all three fluxes.

\subsection*{Historical yield evaluation} 

\begin{figure}[p]
\centering
\makebox[\textwidth][c]{%
\includegraphics[width=1.25\textwidth]{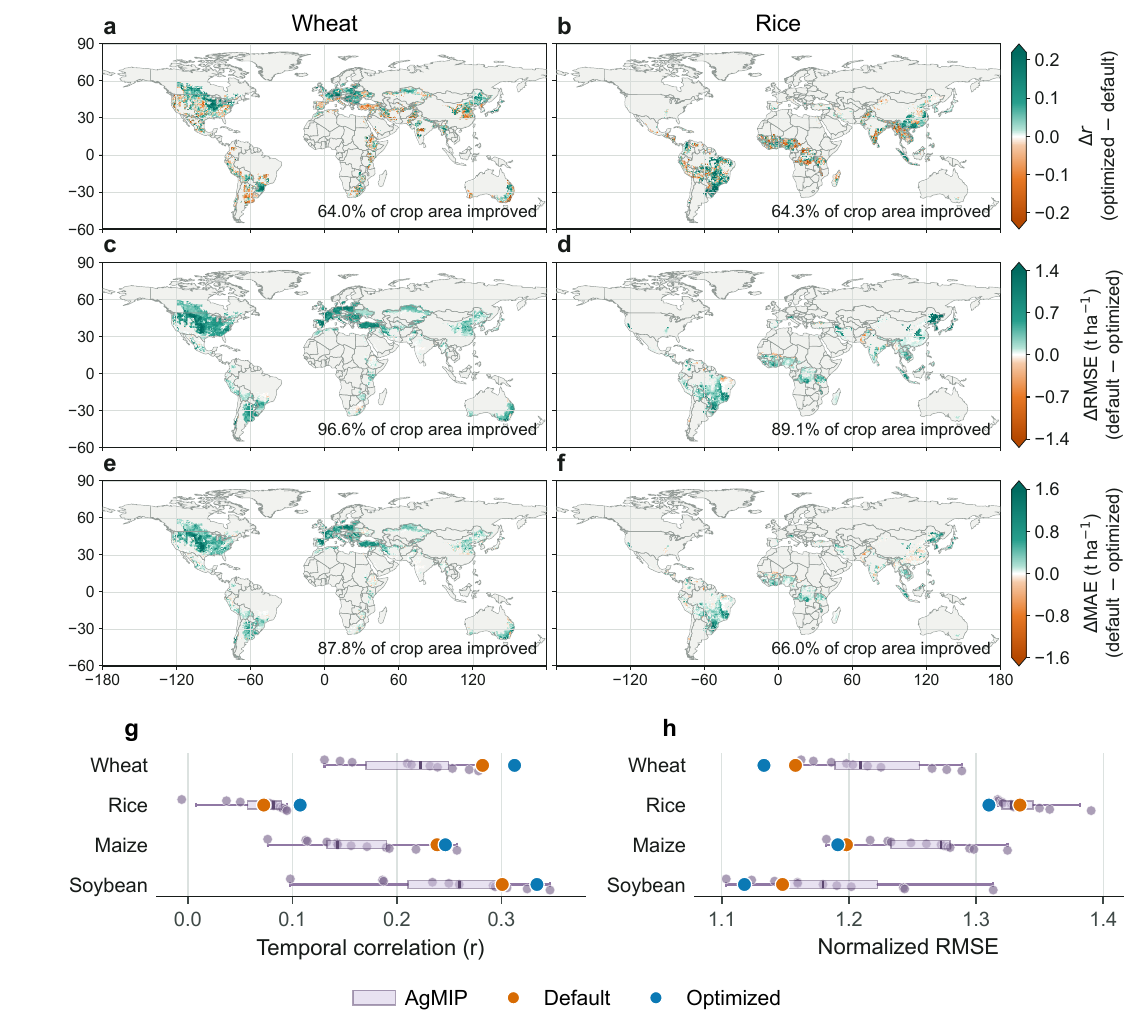}%
}
\caption{\textbf{Observation-based parameter calibration improves historical yield anomaly simulations and low-yield performance.}
\textbf{a, b}, Change in grid-cell temporal correlation between simulations and the global dataset of historical yields (GDHY) \cite{iizumi2020global} for wheat and rice, calculated as $r_{\mathrm{optimized}}-r_{\mathrm{default}}$. 
\textbf{c, d}, Reduction in root-mean-square error (RMSE) for wheat and rice, calculated as $\mathrm{RMSE}_{\mathrm{default}}-\mathrm{RMSE}_{\mathrm{optimized}}$. 
\textbf{e, f}, Spatial changes in mean absolute error (MAE) in years belonging to the lowest 20\% of GDHY anomalies for wheat and rice after parameter optimization. Positive values indicate better performance in all panels. Percentages report the crop area with improved performance. Metrics use grid cells containing at least 10~ha of the crop area and at least five-year yield during 1981--2016. The corresponding maize and soybean maps are shown in Supplementary Fig.~\ref{fig:historical_yield_skill_maize_soybean}. 
\textbf{g--i}, Comparison with GGCMs from AgMIP. Light-purple points show individual GGCMs from AgMIP, boxes show their median and interquartile range, and whiskers extend to 1.5 times the interquartile range. Orange and blue points show the default and optimized Agrocosm simulations. Panels show temporal correlation (\textbf{g}) and normalized RMSE (\textbf{h}). The AgMIP GGCMI includes 13 process-based crop models.}
\label{fig:global_yield_skill}
\end{figure}

Given the strong cross-site performance in fivefold validation, we re-train the selected parameters using flux data for all 100 growing seasons to evaluate their ability to reproduce historical yield variability at the global scale. The final crop and soil parameters are reported in Supplementary Tables~\ref{tab:final_crop_parameterization} and~\ref{tab:final_soil_parameterization}. We run global historical simulations for wheat, rice, maize and soybean under rainfed and irrigated conditions using the default and optimized parameters (\hyperref[sec:global_historical_crop_simulations]{Methods}). We compare the simulations with the global dataset of historical yields (GDHY) for 1981--2016 \cite{iizumi2020global}. To isolate yield interannual variability, simulated and GDHY yield time series are independently detrended at each grid cell (Methods, Eqs.~\ref{eq:gridcell_yield_trend} and~\ref{eq:gridcell_yield_anomaly}).

Although yield data are not used during optimization, grid-cell temporal correlation improves across 64.0\%, 64.3\%, 62.3\% and 61.2\% of the evaluated wheat, rice, maize and soybean area, respectively (Fig.~\ref{fig:global_yield_skill}a,b; Supplementary Fig.~\ref{fig:historical_yield_skill_maize_soybean}a,b). Grid-cell RMSE decreases across 96.6\%, 89.1\%, 67.2\% and 85.1\% of the corresponding evaluated crop areas (Fig.~\ref{fig:global_yield_skill}c,d; Supplementary Fig.~\ref{fig:historical_yield_skill_maize_soybean}c,d). Across the five largest producing countries for each crop, based on mean GDHY production during 1981--2016, temporal correlation increases in 15 of 20 crop--country comparisons and nRMSE decreases in 19 of 20 (Supplementary Table~\ref{tab:top5_country_yield_skill}). The corresponding spatial temporal correlation and RMSE for the default and optimized simulations are shown in Supplementary Figs.~\ref{fig:absolute_yield_correlation} and~\ref{fig:absolute_yield_nrmse}. 

We evaluate the ability of optimized Agrocosm to reproduce low-yield years, which advanced GGCMs often struggle to simulate \cite{schewe2019state, heinicke2022global, auer2026global}. Low-yield years are defined in each grid cell as the lowest 20\% of detrended GDHY anomalies. Relative to the default simulation, the global crop area-weighted nMAE in low-yield years decreases by 43.5\% for wheat, 21.0\% for rice, 3.4\% for maize and 15.4\% for soybean (Fig.~\ref{fig:global_yield_skill}i). Lower grid-cell nMAE occurs across 87.8\%, 66.0\%, 59.2\% and 72.1\% of the corresponding crop areas, respectively (Fig.~\ref{fig:global_yield_skill}e,f; Supplementary Fig.~\ref{fig:historical_yield_skill_maize_soybean}e,f). Using the same GDHY benchmark and evaluation procedure, we further compare Agrocosm with GGCMs that participated in the Agricultural Model Intercomparison and Improvement Project (AgMIP) \cite{franke2020ggcmi,frieler2024scenario}. Based on mean ranks across the four crops, Agrocosm improves from third with default parameters to first with optimized parameters for both temporal correlation and nRMSE (Fig.~\ref{fig:global_yield_skill}g, h). The optimized simulations also have lower nMAE in low-yield years than the AgMIP ensemble median for all four crops (Supplementary Fig.~\ref{fig:historical_low_yield_nmae_benchmark}).

We also examine how the site-trained parameters change the agreement between simulated seasonal processes and reported yield variability over 1981--2016. The temporal correlation between aboveground-biomass anomalies and GDHY yield anomalies increases for all four crops. Correlation with seasonal ET also increases for every crop, while correlation with GPP increases for wheat, rice and maize (Supplementary Fig.~\ref{fig:process_yield_correlations}). These results indicate that simulated biomass and seasonal ET are more strongly correlated with observed yield anomalies after flux-based parameter optimization.

\subsection*{Future projections across parameterizations}

\begin{figure}[p]
\centering
\makebox[\textwidth][c]{
\includegraphics[width=1.05\textwidth]{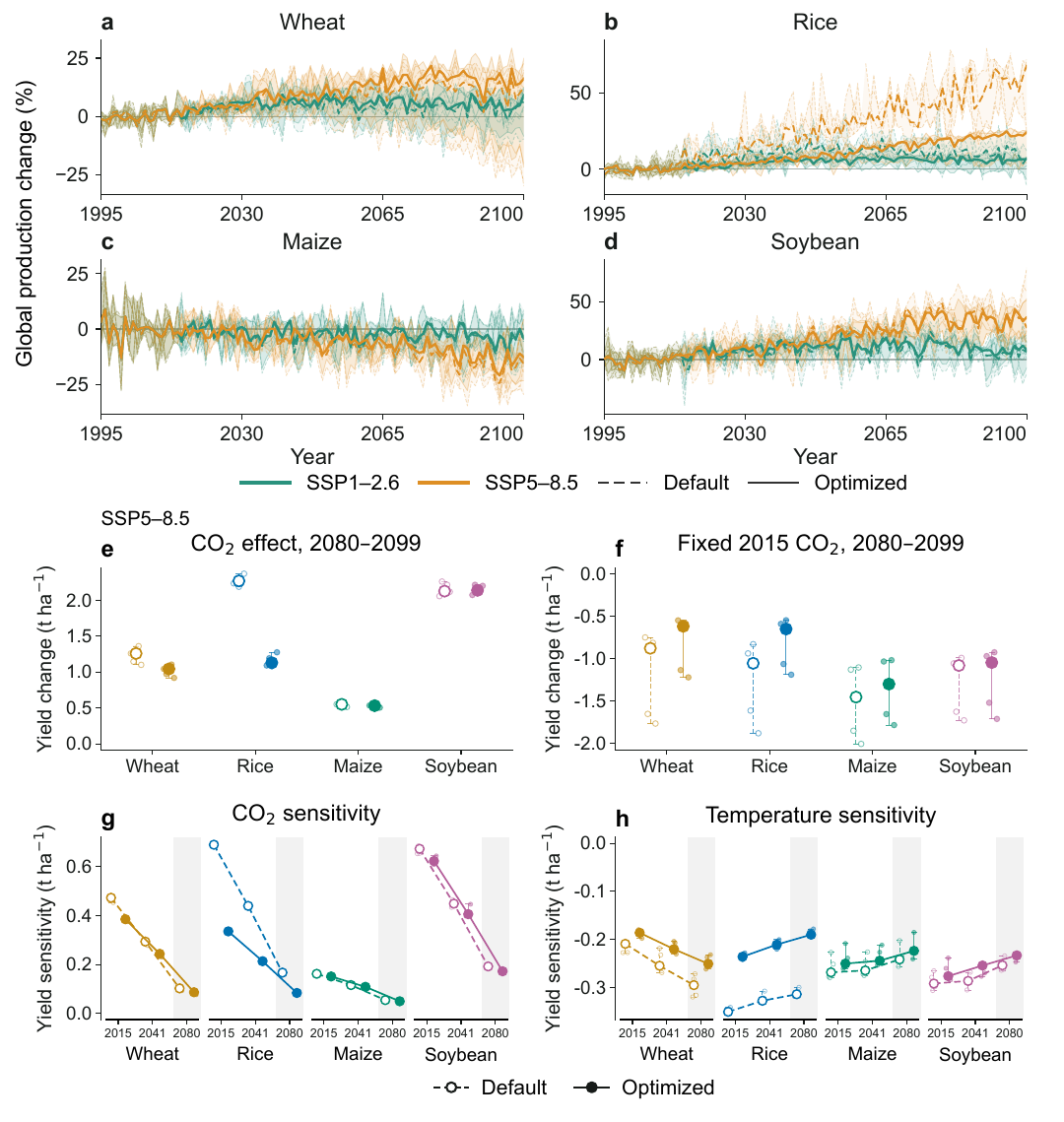}
}
\caption{\textbf{Parameter optimization changes future global crop production responses and hotspots of strong yield decline or increase.}
\textbf{a--d}, Projected global production changes for wheat, rice, maize and soybean under SSP1--2.6 and SSP5--8.5 across five climate models, i.e., GFDL-ESM4, IPSL-CM6A-LR, MPI-ESM1-2-HR, MRI-ESM2-0 and UKESM1-0-LL. Dashed and solid lines denote the default and optimized parameterizations, respectively. Shading shows the corresponding interquartile range. Production in each parameterization and climate model is expressed relative to its own 1995--2014 mean. 
\textbf{e--h}, Yield responses and climate sensitivities under SSP5--8.5.
\textbf{e}, CO$_2$ effect, calculated as the difference in mean yield over 2080--2099 between simulations with rising CO$_2$ and CO$_2$ fixed at its 2015 concentration (399.95 ppm).
\textbf{f}, Yield change over 2080--2099 relative to the corresponding 1995--2014 mean with CO$_2$ fixed at its 2015 concentration. All other climate inputs follow SSP5--8.5.
\textbf{g,h}, Yield sensitivities to CO$_2$ concentration and temperature, respectively, calculated by reverse-mode automatic differentiation and scaled to 100 ppm and 1\,$^\circ$C. Values are averaged over 2015--2034, 2041--2060 and 2080--2099. Ticks mark the starting years and grey bands highlight the last period. Small points show individual climate models. Large points and whiskers show their median and full range. Dashed lines and open points denote default parameters. Solid lines and filled points denote optimized parameters.}
\label{fig:future_yield_response}
\end{figure}

\begin{figure}[p]
\centering
\makebox[\textwidth][c]
{\includegraphics[width=1.05\textwidth]{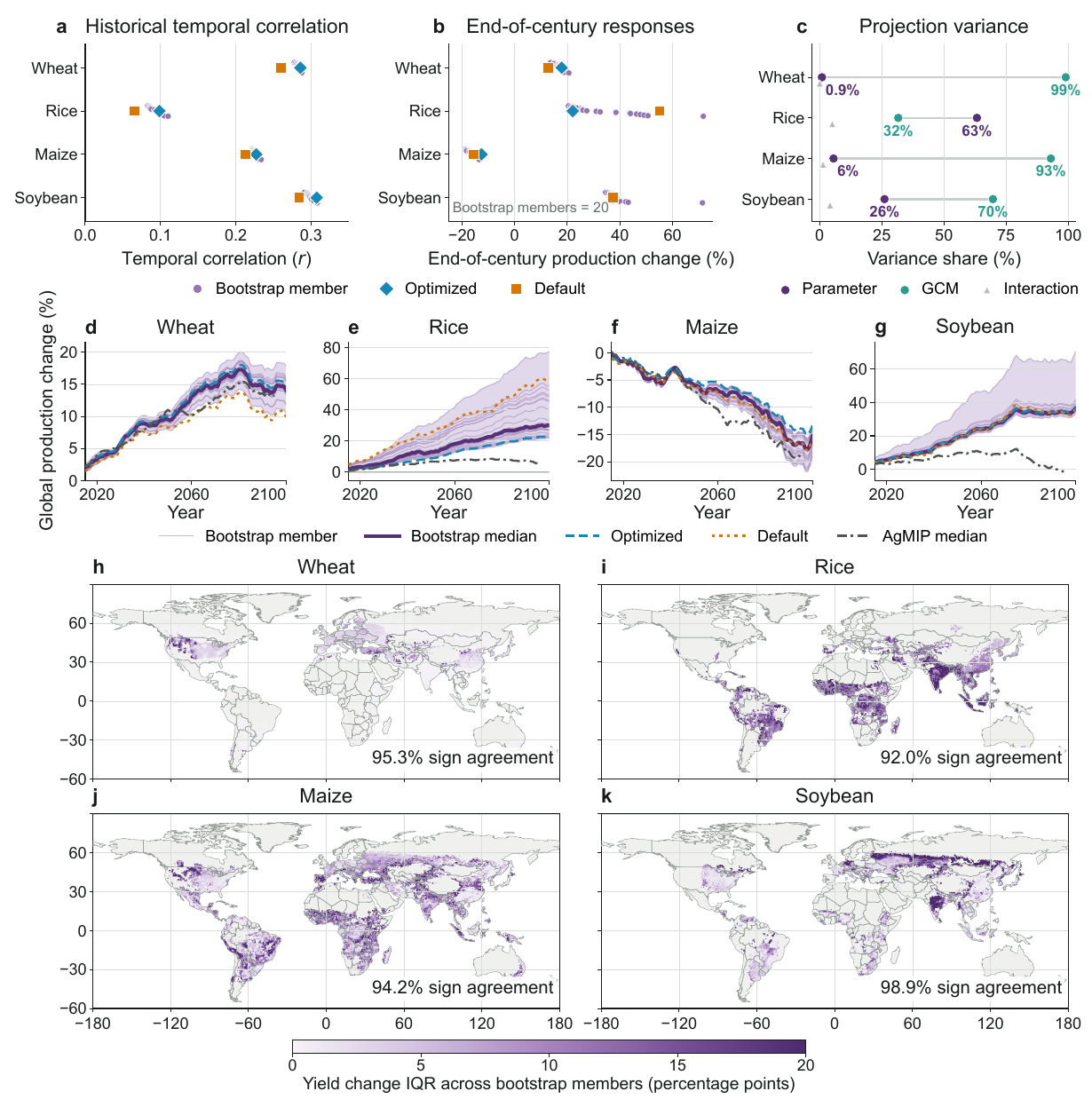}
}
\caption{\textbf{Parameterizations with similar historical performance produce different future projections.}
Twenty parameterizations per crop are compared under SSP5--8.5, obtained by recalibration on bootstrap resamples of flux-tower sites. Global production and grid-cell yield changes are expressed relative to the corresponding 1995--2014 mean for each parameterization and climate model.
\textbf{a}, Historical temporal correlation ($r$) against the global dataset of historical yields (GDHY) for 1981--2016 \cite{iizumi2020global}.
\textbf{b}, End-of-century (2080--2099) global production changes summarized by the five-model median. In \textbf{a, b}, purple points show individual bootstrap members. Blue diamonds and orange squares show the optimized parameterization calibrated using all sites and the default parameterization, respectively.
\textbf{c}, Variance shares in end-of-century global production changes attributable to parameters, climate models and their interaction.
\textbf{d--g}, Global production trajectories. Thin purple lines show the five-model medians of individual bootstrap members. The thick purple line shows the ensemble median, and shading spans the range of the smoothed bootstrap-member curves. Blue and orange dash lines show the optimized and default parameterizations. The grey line shows the AgMIP ensemble median. Curves use centered 11-year moving averages.
\textbf{h--k}, Interquartile range ($Q_{75}-Q_{25}$) of grid-cell yield changes for 2080--2099 relative to 1995--2014 across bootstrap members, calculated from their five-model median responses and expressed in percentage points. Values above 20 percentage points use the upper colour limit. Annotations report the percentage of crop area where at least 17 of the 20 bootstrap members agree on the sign of yield change.}
\label{fig:parameter_uncertainty}
\end{figure}

We compare global simulations using the default and optimized parameter sets under SSP1--2.6 and SSP5--8.5, with climate forcing from five climate models and atmospheric CO$_2$ concentrations prescribed according to the corresponding SSP (\hyperref[sec:future_crop_simulations]{Methods}). Global production change is expressed relative to the 1995--2014 mean. Under SSP5--8.5, optimization changes end-of-century (2080--2099) production response from 12.8\% to 17.9\% for wheat, 55.1\% to 22.1\% for rice and $-15.5$\% to $-12.5$\% for maize, while soybean remains nearly unchanged at 37.4\% (Fig.~\ref{fig:future_yield_response}a--d). The same directions of change occur under SSP1--2.6, although the differences are smaller at $+0.8$, $-6.1$, $+0.8$ and $+0.1$ percentage points for wheat, rice, maize and soybean, respectively. These results show that the parameter optimization with respect to fluxnet observations changes crop-specific responses to future climate, with larger effects under SSP5--8.5. The effects of the five individual climate models are reported in Supplementary Table~\ref{tab:climate_model_parameterization_effects}. The end-of-century grid-cell patterns under SSP1--2.6 and SSP5--8.5 are shown in Supplementary Figs.~\ref{fig:future_yield_maps_end_of_century_ssp126} and~\ref{fig:future_yield_maps_end_of_century_ssp585}, respectively.

We also compare simulations with CO$_2$ following SSP5--8.5 or held fixed at its 2015 concentration. The latter projections show that end-of-century yield changes are negative for all four crops (Fig.~\ref{fig:future_yield_response}f). Optimization most strongly reduces the CO$_2$ benefit for rice, from 2.27 to 1.13~t~ha$^{-1}$ (Fig.~\ref{fig:future_yield_response}e), while its fixed-CO$_2$ yield change becomes less negative, from $-1.06$ to $-0.65$~t~ha$^{-1}$. Automatic differentiation further reveals contrasting changes in CO$_2$ and temperature sensitivities. CO$_2$ sensitivity declines across the three periods for all four crops under both parameterizations (Fig.~\ref{fig:future_yield_response}g). For rice, optimization approximately halves this sensitivity in each period, consistent with the reduced CO$_2$ benefit. Temperature sensitivities remain negative but become more negative over time for wheat and less negative for rice, maize and soybean (Fig.~\ref{fig:future_yield_response}h). Optimization makes temperature sensitivities less negative for all four crops, with the largest end-of-century change in rice from $-0.31$ to $-0.19$~t~ha$^{-1}$ per $^\circ$C.

To quantify the effect of parameter uncertainty, we repeat parameter optimization for 20 bootstrap resamples of the 27 flux-tower sites and propagate the resulting parameter sets through future simulations under two scenarios and five climate models (\hyperref[sec:parameter_ensemble]{Methods}). All 20 bootstrap members have higher temporal correlation with GDHY than the default simulations and cluster closely around the corresponding optimized simulation (Fig.~\ref{fig:parameter_uncertainty}a). Yet their end-of-century production responses under SSP5--8.5 span 15.09--18.78\%, 21.83--49.35\%, $-18.47$\% to $-13.49$\% and 36.15--42.30\% for wheat, rice, maize and soybean, respectively (Fig.~\ref{fig:parameter_uncertainty}b,d--g; Supplementary Table~\ref{tab:four_crop_parameter_ensemble_skill}). Comparable historical skill therefore does not imply a unique end-of-century response. Variance decomposition attributes 63\% of the crossed-ensemble variance in rice production responses to parameter sampling, compared with 0.9\%, 6\% and 26\% for wheat, maize and soybean, respectively (Fig.~\ref{fig:parameter_uncertainty}c; Supplementary Table~\ref{tab:parameter_gcm_variance}).

The spatial maps reveal pronounced heterogeneity in parameter uncertainty (Fig.~\ref{fig:parameter_uncertainty}h--k). At least 17 of 20 parameter sets agree on response direction across 92.0--98.9\% of crop area, but the magnitude of the spread varies among regions and crops. At each grid cell, we quantify this spread as the range covered by the middle 50\% (interquartile range IQR) of projected yield changes from the 20 parameter sets. A yield-change IQR of at least 10\% occurs across 3.2\%, 74.3\%, 27.4\% and 13.2\% of the wheat, rice, maize and soybean area. High-IQR wheat regions cluster in the United States and western Asia, maize in Brazil and sub-Saharan Africa, and soybean in India and northern Eurasia. For rice, parameter uncertainty is both larger and more spatially extensive. Its crop area-weighted median IQR is 16.14\%, and 74.3\% of its area exceeds 10\%. Both crop production responses and parameter uncertainty are generally smaller under SSP1--2.6 (Supplementary Fig.~\ref{fig:parameter_uncertainty_ssp126}).

\subsection*{Extreme event attribution}

\begin{figure}[p]
\centering
\makebox[\textwidth][c]{\includegraphics[width=1.0\textwidth]{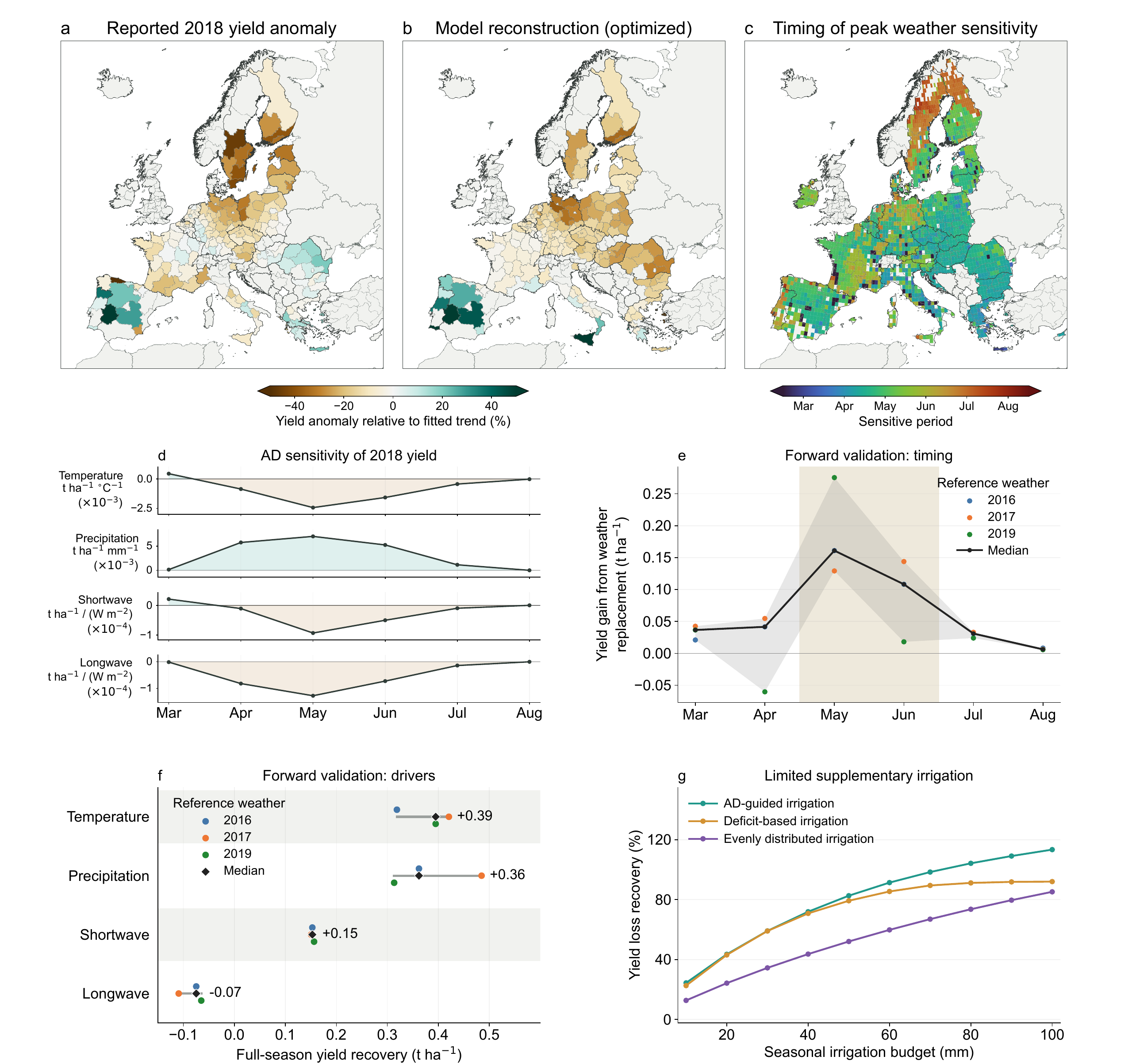}}
\caption{\textbf{Differentiability attribute weather conditions to the 2018 European wheat yield anomaly.}
\textbf{a}, Reported 2018 yield anomalies from the harmonized European Union subnational crop statistics \cite{ronchetti2023harmonized}. 
\textbf{b}, Simulated yield anomalies aggregated to the same administrative units. Anomalies are calculated as deviations from the corresponding fitted trends.
\textbf{c}, Timing of peak weather sensitivity in each grid cell. Colours indicate the timing of the 14-day period with the highest cumulative weather sensitivity in each grid cell. Daily AD sensitivities are scaled by the standard deviation of each weather variable before their absolute values are summed.
\textbf{d}, Monthly mean daily AD sensitivities of simulated end-of-season yield to each weather variable, averaged across cells with crop-area weights. Positive and negative values indicate whether an increase in the weather input raises or lowers end-of-season yield.
\textbf{e, f}, Yield changes relative to the original 2018 simulation using weather from 2016, 2017 or 2019. For \textbf{e}, a 14-day window is moved through the growing season. Each simulation replaces all four weather inputs only within one window, while other days retain 2018 weather. Responses are averaged by month. For \textbf{f}, one weather variable is replaced throughout the post-sowing season, with all other inputs unchanged. Coloured points denote reference years. Black lines (\textbf{e}) and diamonds (\textbf{f}) show medians. Shading and horizontal lines show full ranges, respectively.
\textbf{g}, Yield recovery under AD-guided (green), deficit-based (orange) and evenly distributed (purple) irrigation with seasonal budget limits of 10--100~mm. Recovery is the irrigation-induced yield gain divided by the gap between fitted-trend and 2018 rainfed yields. AD-guided irrigation use AD to determine the irrigation volume. Deficit-based irrigation sets the daily irrigation to the crop water deficit. Evenly distributed irrigation allocates the full seasonal budget distributed evenly across growing days. Actual water use under the AD-guided and deficit-based schemes can be lower than the budget. Values above 100\% indicate yields exceeding the fitted trend. Panels \textbf{d--f} focus on the quarter of crop area with the most negative reported yield anomalies.}
\label{fig:extreme_event_attribution}
\end{figure}

We finally use differentiability to diagnose weather contributions to a yield-loss event. European wheat in 2018 provides a well-documented example, with widespread crop yield losses across northern and eastern Europe \cite{beillouin2020impact}. Simulations with optimized parameters captures the broad spatial pattern of 2018 yield anomalies compared with the harmonized European Union subnational crop statistics \cite{ronchetti2023harmonized}, with a crop area-weighted spatial correlation of 0.652 (Fig.~\ref{fig:extreme_event_attribution}a,b). Anomalies are calculated relative to trends fitted separately to the observed and simulated yield time series. Within the quarter of crop area most affected by observed yield losses, reported and simulated yields are 25.0\% and 20.6\% below trend respectively, with agreement in yield change direction across 98.2\% of the area.

Using reverse-mode AD, we calculate the sensitivity of end-of-season yield to daily weather inputs along the simulated 2018 crop trajectory (\hyperref[sec:yield_loss_attribution]{Methods}). At each grid cell, we identify the 14-day period with the largest combined sensitivity to temperature, precipitation, shortwave radiation and longwave radiation (Fig.~\ref{fig:extreme_event_attribution}c, see also \hyperref[sec:yield_loss_attribution]{Methods}). Within the worst-affected quarter of crop area, monthly mean sensitivities indicate that small increases in rainfall would raise yield, whereas increases in temperature, shortwave and longwave radiation would lower yield during April--June. Monthly mean sensitivities to all four weather variables are strongest in May (Fig.~\ref{fig:extreme_event_attribution}d). The temperature and precipitation sensitivities are consistent with evidence linking the 2018 European wheat yield losses to heat and drought \cite{beillouin2020impact}. We then use counterfactual forward simulations to evaluate whether the months of high weather sensitivity identified by AD also show large yield responses to weather changes. In each simulation, we replace all four weather inputs within a single 14-day window with the corresponding inputs from a reference year (2016, 2017 or 2019), with all other days unchanged.We shift this window through the growing season and repeat the simulation at each position. The largest median yield gain occurs in May ($+0.16$~t~ha$^{-1}$), followed by June ($+0.11$~t~ha$^{-1}$; Fig.~\ref{fig:extreme_event_attribution}e), consistent with the May peak in AD sensitivities. Across the three reference years, replacing temperature or precipitation alone throughout the post-sowing season produces median yield gains of $+0.39$ and $+0.36$~t~ha$^{-1}$, respectively (Fig.~\ref{fig:extreme_event_attribution}f).

We next demonstrate how AD enables more effective irrigation strategies. We compare AD-guided, deficit-based and evenly distributed supplementary irrigation for alleviating the simulated 2018 yield loss. The AD-guided scheme uses AD to quantify how the daily crop water deficit responds to additional irrigation and determines the irrigation volume from this sensitivity. The deficit-based scheme instead sets the daily irrigation to the crop water deficit. Both schemes make daily decisions using only current crop states and weather, subject to the same seasonal water-budget limit as the uniform scheme. Across the predefined 10--100~mm range, AD-guided irrigation recovers 24.4--113.4\% of the crop area-weighted simulated yield deficit relative to the fitted trend, compared with 22.6--92.0\% under deficit-based irrigation and 12.7--85.2\% under uniform application (Fig.~\ref{fig:extreme_event_attribution}f). As the seasonal budget increases from 50 to 100~mm, yield recovery under deficit-based irrigation approaches a plateau, while recovery under AD-guided irrigation continues to rise. Its recovery advantage ranges from 3.4 to 21.4 percentage points, highlighting the potential of AD to achieve additional yield gains beyond deficit-based irrigation.

\section*{Discussion}

Our results demonstrate the advantages of systematic gradient-based parameter optimization in a fully process-based GGCM. The generalization from site-level flux observations to independent global yield variability shows that calibration improves model behaviour beyond the observations used for optimization. Compared with 13 AgMIP GGCMs \cite{franke2020ggcmi}, optimization moves Agrocosm upward from its default position, making it highly competitive across crops. The AgMIP comparison shows that no GGCM performs consistently well across all crops. We argue that extending differentiable calibration across these GGCMs could improve individual model performance and thereby strengthen the ensemble considerably.

Parameter optimization not only improves historical performance but also reshapes projected crop responses to future climate. For rice under SSP5--8.5, optimization reduces the projected end-of-century production increase from 55.1\% to 22.1\% (Fig.~\ref{fig:future_yield_response}b). Automatic differentiation further indicates contrasting changes in CO$_2$ and temperature sensitivities for all four crops. The bootstrap ensemble further reveals that parameterizations with similar historical skill can produce markedly different future responses, with parameter-driven uncertainty varying among crops and regions. For rice, our results show that parameter uncertainty accounts for 28.1\% of projection variance under SSP1--2.6 and rises to 63.3\% under SSP5--8.5, exceeding the 31.6\% attributed to the differences in driving climate models (Fig.~\ref{fig:parameter_uncertainty}c; Supplementary Fig.~\ref{fig:parameter_uncertainty_ssp126}c). Parameter uncertainty is also substantial for soybean but smaller for maize and wheat. Current global crop-model intercomparisons quantify uncertainty across crop and climate models but typically leave within-model parameter uncertainty unaccounted for \cite{wallach2025there}. This suggests that the food-security assessments based on a single parameterization may underestimate uncertainty in crop projections under future climate change.

The 2018 European wheat analysis shows that differentiability can go beyond improved simulations of yields and other variables, but can also diagnose the weather forcing and within-season periods that determine these variables. Conventional GGCM sensitivity analyses require large ensembles of forward simulations under prespecified perturbations \cite{franke2020ggcmi,franke2020emulators}. AD instead calculates the sensitivity of end-of-season yield to daily weather inputs directly from simulated process trajectories. In our case, the analysis highlights May temperature and June precipitation as the leading contributors to the simulated yield loss. This diagnosis is consistent with independent evidence linking the 2018 European wheat losses to heat and drought \cite{beillouin2020impact, makinen2018sensitivity}, supporting the credibility of this model-based diagnosis. The irrigation results further demonstrate how AD enables more effective irrigation strategies. AD-guided irrigation alleviates simulated yield losses during extreme heat and drought more effectively than deficit-based and evenly distributed irrigation. Yield recovery under AD-guided irrigation continues to rise even as gains from deficit-based irrigation approach a plateau.

Despite these advances, several challenges remain in our study. Flux-tower observations are limited and unevenly distributed (Supplementary Fig.~\ref{fig:flux_site_distribution}), limiting robust estimation of spatially varying parameters. Our global simulations hence use a single parameter set for each crop, although physiological thresholds may vary among regions~\cite{wang2026spatially}. Some parameters approach their prescribed bounds during calibration, which may indicate process compensation \cite{seidel2018towards}. Additional observations of phenological development would improve calibration, but heterogeneity among existing datasets currently limits their direct use in calibrating process-based models \cite{noia2025crop}. The 20-member bootstrap quantifies uncertainty across alternative parameterizations, but does not cover the full range of parameters. Differentiable attribution is limited to events that the model can reproduce and represent. Effects of unrepresented management, pests or diseases therefore remain unresolved. 

The modest improvements in low-yield simulations for rice, maize and soybean further show that accurately representing yield failure due to heat, drought and excess water remains a major challenge \cite{schewe2019state, heinicke2022global, auer2026global}. The differentiable modelling framework could help address this challenge, but doing so requires improved stress representations, harmonized observations and independent validations. Satellite data may provide broader spatial constraints once their scales and uncertainties are reconciled with model variables. Our framework can also support the optimization of management practices for specific objectives to strengthen assessments of future food security. We believe that the differentiable framework demonstrated here has the potential to advance process-based modelling across a wide range of scientific applications.

\section*{Methods}

\subsection*{Model description}

The process-based global gridded crop model (GGCM) used in this study builds on crop and soil process formulations and default parameters adapted from LPJmL ("Lund--Potsdam--Jena managed Land")~\cite{bondeau2007modelling, 2018LPJmL, von2018implementing, lutz2019simulating, wirth2024biological}. Independently implemented in Julia as Agrocosm, the model simulates coupled carbon, nitrogen, water and energy dynamics in crop--soil systems at a daily time step. The implementation features a redesigned CPU/GPU-compatible numerical architecture, a distinct soil initialization and agricultural spin-up procedure (Supplementary Information section~\ref{sec:hwsd_spinup}), and support for automatic differentiation, a capability not natively available in LPJmL. The spatial resolution of global simulations is 0.5$^{\circ}\times0.5^{\circ}$ (latitude × longitude) in this study.

Agrocosm provides a separate implementation of daily crop and soil calculations for automatic differentiation. This enables gradient-based optimization of continuous crop and soil parameters through growing-season simulations without modifying the standard forward-simulation routines. In this study, daily gross primary production (GPP), ecosystem respiration (RECO) and total evapotranspiration (ET) are jointly used to constrain crop and soil parameters. Agrocosm’s backend-independent numerical design supports execution on CPUs and GPUs across floating-point types, providing a scalable basis for large-scale parameter optimization.

\subsection*{Model spin-up} %
Unlike the natural-vegetation spin-up used by LPJmL, Agrocosm initializes agricultural soil states from the Harmonized World Soil Database (HWSD) version 2.0 \cite{fao2023hwsd}. HWSD provides static soil properties and layer-resolved targets for soil organic carbon (SOC) and total nitrogen, which are converted to areal stocks and remapped to the Agrocosm soil layers. Layer remapping and deep soil extrapolation are described in the Supplementary Information (section~\ref{sec:hwsd_mapping}).

We perform a 600-year target-constrained agricultural spin-up using fixed 2015 crop management \cite{frieler2024scenario} and a repeating 1901--1930 daily climate sequence. During spin-up, simulated SOC and total nitrogen are constrained to the HWSD-derived targets while litter pools evolve freely. The resulting soil-pool partitioning and static soil properties are used to initialize the crop simulations (Supplementary Information, sections~\ref{sec:agricultural_spin} and~\ref{sec:initial_soil_state}). Static soil properties are used to derive hydraulic and thermal parameters. Initial soil water in each layer is set to field capacity calculated using pedotransfer functions. The spin-up and state initialization
remain outside the differentiable daily transition.

\subsection*{Gradient-based model optimization}
\label{sec:gradient-based_model_optimization}

Differentiable programming uses automatic differentiation (AD) to calculate gradients through numerical programs. \cite{gelbrecht2023differentiable}. Using the chain rule, AD calculates how model outputs respond to changes in model parameters or continuous inputs \cite{baydin2018automatic}. We formulate the daily Agrocosm simulation as a parameterized state-space transition,
\begin{equation}
\mathbf{x}_{t}
=
F\left(
\mathbf{x}_{t-1},
\mathbf{u}_{t},
\mathbf{m}_{t};
\boldsymbol{\theta}
\right),
\end{equation}
\begin{equation}
\mathbf{y}_{t}
=
H\left(
\mathbf{x}_{t},
\mathbf{u}_{t},
\mathbf{m}_{t};
\boldsymbol{\theta}
\right),
\end{equation}
where $\mathbf{x}_{t}$ is the prognostic crop--soil state, $\mathbf{u}_{t}$ is the daily climate forcing, $\mathbf{m}_{t}$ is the management input, $\boldsymbol{\theta}$ is the selected vector of continuous model parameters, and $\mathbf{y}_{t}$ contains the diagnostic fluxes used for calibration. The initial state $\mathbf{x}_{0}$ is constructed before differentiation from the HWSD-based soil initialization and agricultural spin-up described above. 

We use Enzyme.jl \cite{moses2021enzyme} to apply reverse-mode automatic differentiation through the selected continuous state transitions and compute gradients of the observation loss with respect to the continuous model parameters over the full daily trajectory. The spin-up, state initialization and output writing remain outside the differentiable trajectory. Discrete events, including sowing, harvest and crop-failure decisions, are resolved before differentiation and are treated as fixed during gradient evaluation. The resulting derivatives therefore describe the selected continuous simulation path rather than assuming that every branch of the complete crop model is globally differentiable.

In this study, we calibrate Agrocosm against daily GPP, RECO and ET from agricultural flux sites in the FLUXNET products \cite{pastorello2020fluxnet2015}. The calibration dataset contains 100 growing-season records from 27 physical sites, described in detail in the Supplementary Information section~\ref{sec:fluxnet_data}. To select parameters that affect simulated GPP, RECO and ET and can be constrained by the observations, we assess the sensitivity of these fluxes to each candidate parameter and the change in fit when each is excluded in turn (Supplementary Information section~\ref{sec:parameter_selection}). The final selection contains 12 crop-specific parameters and four soil parameters with prescribed lower and upper bounds (see Supplementary Tables~\ref{tab:cft_parameters} and~\ref{tab:soil_parameters}).

We calibrate crop and soil parameters in two stages. We first optimize the crop parameters against GPP, RECO and ET, then optimize soil parameters against RECO and ET while holding the calibrated crop parameters fixed. (Supplementary Information section~\ref{sec:two_stage_calibration}). This design is intended to limit compensation between their effects and allows us to assess the additional improvement from soil parameter calibration. Grouped-site fivefold cross-validation is first used to assess cross-site model generalization, with all seasons and crops from the same physical site assigned to the same fold. We then re-train the selected parameters using all observational data for global simulations. The gradient-based optimizer Adam \cite{kingma2015adam} is used to update the parameters at each calibration iteration. Every Adam update is clipped to remain within these bounds.

\subsection*{Historical crop simulations} %
\label{sec:global_historical_crop_simulations}
We perform daily global simulations for wheat (including winter and spring wheat), rice, maize and soybean on the 0.5$^{\circ}\times0.5^{\circ}$ grid. For each crop, we conduct a baseline simulation using the default parameters and a simulation using the crop-specific and soil parameters optimized against daily observations of GPP, RECO and ET. The site-trained crop and soil parameter products are reported in Supplementary tables~\ref{tab:final_crop_parameterization} and~\ref{tab:final_soil_parameterization}. Historical climate forcing is taken from the ISIMIP3a GSWP3-W5E5 climate input, including near-surface air temperature, precipitation, and shortwave and net longwave radiation, and atmospheric CO$_2$ \cite{frieler2024scenario}. Each crop is simulated separately under rainfed and fully irrigated conditions. Each crop is simulated separately under rainfed and idealized fully irrigated conditions, following the GGCMI Phase 3 setup \cite{jagermeyr2021climate}. In the irrigated simulations, soil water is restored to field capacity each day without a limit on water supply. Simulations start from the initial soil state taken from the spin-up and run continuously from 1901 to 2019 using a 365-day calendar. All crop management inputs, including land-use fractions, sowing dates, phenological heat requirements, fertilizer and residue management, are taken from the ISIMIP3a and held at their 2015 levels following ISIMIP3a protocol.

We evaluate the simulated total yields against the global dataset of historical yields (GDHYv1.2+v1.3), which provides 0.5$^{\circ}\times0.5^{\circ}$ gridded total yields for wheat, maize, rice and soybean from 1981 to 2016 \cite{iizumi2020global}. GDHY combines national yield statistics from FAO with satellite-derived vegetation data, including leaf area index and the fraction of absorbed photosynthetically active radiation, to estimate grid-cell yields~\cite{iizumi2020global}. We exclude grid cells with crop area less than $10\text{ ha}$, a threshold below which we consider crop production to be too marginal to be included in the analysis. We detrend the yield time series of GDHY to remove long-term trends associated with technological progress, including improved cultivars and management practices, and thereby better isolate climate-related interannual yield variability. For each grid cell, the simulated and GDHY yield time series are detrended independently for all four crops. For $d\in\{\mathrm{default},\mathrm{optimized},\mathrm{GDHY}\}$, we fit
\begin{equation}
\label{eq:gridcell_yield_trend}
Y^{d}_{g,c,t}
=
a^{d}_{g,c}
+
b^{d}_{g,c}\tau_t
+
\varepsilon^{d}_{g,c,t},
\end{equation}
Here $Y^d_{g,c,t}$ is annual yield for grid cell $g$, crop $c$ and year $t$, with $d$ denoting the default simulation, optimized simulation or GDHY. The terms $a^d_{g,c}$ and $b^d_{g,c}$ are the intercept and slope of the linear trend, $\tau_t$ is the centred calendar year, and $\varepsilon^d_{g,c,t}$ is the residual. Using the fitted coefficients $\widehat{a}^d_{g,c}$ and $\widehat{b}^d_{g,c}$, we calculate the detrended yield anomaly as
\begin{equation}
\label{eq:gridcell_yield_anomaly}
\widetilde{Y}^{d}_{g,c,t}
=
Y^{d}_{g,c,t}
-
\widehat{a}^{d}_{g,c}
-
\widehat{b}^{d}_{g,c}\tau_t .
\end{equation}

We additionally assess the ability of Agrocosm with optimized parameters to reproduce low-yield conditions. Low-yield years are defined in each grid cell as the lowest 20\% of detrended GDHY yield anomalies. For these years, we calculate the mean magnitude of the difference between the detrended simulated and GDHY anomalies and normalize it by the temporal standard deviation of the corresponding GDHY anomaly series. To diagnose associated seasonal processes, we calculate grid-cell temporal correlations between GDHY yield anomalies and seasonal GPP, LAI-days, ET, water deficit, growing-season length and aboveground biomass. The evaluation metrics are described in Supplementary Information section~\ref{sec:evaluation_metrics}.

For the multi-model comparison, we evaluate historical simulations from AgMIP GGCMs using the same GDHY series, period, detrending, metrics and crop area weights \cite{franke2020ggcmi,frieler2024scenario}. AgMIP includes 13 GGCMs, i.e., ACEA \cite{mialyk2024water}, CROVER~\cite{okada2015modeling}, EPIC-IIASA \cite{balkovivc2014global}, ISAM~\cite{song2013implementation}, LDNDC~\cite{haas2013landscapedndc}, LPJmL~\cite{von2018implementing}, LPJ-GUESS \cite{ma2022modeling}, PEPIC \cite{williams1989epic}, DSSAT-Pythia \cite{joshi2026pythia}, pDSSAT~\cite{jones2003dssat}, PROMET~\cite{mauser2015global}, CYGMA1p74 \cite{iizumi2017responses} and SIMPLACE-LINTUL5~\cite{enders2023simplace}.

\subsection*{Future crop simulations} 
\label{sec:future_crop_simulations}
For the future projections, we also perform simulations with the default and optimized parameter sets for wheat (winter and spring wheat), rice, maize and soybean. Each projection includes a historical simulation for model initialization, followed by future simulations under SSP1--2.6 and SSP5--8.5. The two parameterizations use identical climate forcing and crop management inputs. Daily climate forcing is taken from the bias-adjusted ISIMIP3b \cite{frieler2026scenario}, which includes five CMIP6 Earth system models: GFDL-ESM4, IPSL-CM6A-LR, MPI-ESM1-2-HR, MRI-ESM2-0 and UKESM1-0-LL. Atmospheric CO$_2$ concentrations follow the prescribed annual trajectory for each SSP and are identical across climate models and parameterizations within each scenario. Historical simulations cover 1901--2014 and future simulations cover 2015--2100. Management inputs remain fixed at their 2015 values. Rainfed and fully irrigated systems are simulated separately.

For each parameterization and climate model, we calculate annual grid-cell yield changes relative to the corresponding 1995--2014 mean. Following Jägermeyr et al.~\cite{jagermeyr2021climate}, we aggregate these changes globally using observed production weights based on fixed-2015 crop areas and SPAM reference yields \cite{ifpri2019spam2010}. End-of-century production responses are calculated by averaging the annual responses over 2080--2099. We repeat the 2015--2100 simulations with atmospheric CO$_2$ fixed at its 2015 concentration (399.95 ppm). The CO$_2$ effect is calculated as the difference in mean yield over 2080--2099 between the rising-CO$_2$ and fixed-CO$_2$ simulations.

We calculate yield sensitivities to atmospheric CO$_2$ concentration and daily temperature using reverse-mode automatic differentiation. For each growing season, daily derivatives of harvest yield are summed from after sowing to before harvest and scaled to 100 ppm or 1\,$^\circ$C, respectively. These sensitivities provide an estimate of how yield would change if CO$_2$ were 100 ppm higher or temperature were 1\,$^\circ$C warmer throughout the growing season. Seasonal sensitivities are summed across harvests within each year and aggregated globally using weights based on fixed-2015 crop areas and SPAM reference yields. Annual sensitivities are averaged over 2015--2034, 2041--2060 and 2080--2099 for each climate model, then summarized by the median and full range across the five models.

\subsection*{Parameter uncertainty}
\label{sec:parameter_ensemble}
We assess the sensitivity of calibrated parameters to the choice of flux-tower sites using a site-level bootstrap. For each of 20 replicates, we draw 27 physical sites with replacement from the original 27, and repeat the crop and soil parameter calibration with unchanged settings. This ensemble captures variation from resampling the available sites. For each crop, we run simulations using all 20 bootstrap-derived parameter sets under SSP1--2.6 and SSP5--8.5 across GFDL-ESM4, IPSL-CM6A-LR, MPI-ESM1-2-HR, MRI-ESM2-0 and UKESM1-0-LL. We evaluate each parameter set against GDHY using historical simulations driven by the climate forcing GSWP3-W5E5 and the same metrics used for the parameterization trained on all sites. Future production changes are calculated relative to the 1995--2014 mean of the corresponding simulation.

We analyse the parameter-ensemble projections at global and grid-cell scales. At the global scale, we calculate annual production responses separately for each parameter set and climate model. We then take the median across climate models for each parameter set and year. To characterize the spread in end-of-century global production responses across the $20\times5$ simulations, we perform a two-factor variance decomposition using bootstrap parameter set and driving climate model as the two factors. Their interaction captures whether differences between parameter sets vary across climate models. We report the fraction of total variance associated with each factor and their interaction. We repeat the calculation after omitting each member for every crop.

At the grid-cell scale, we first calculate end-of-century yield responses separately for each parameter set and climate model, then take the five-model median at each cell for each parameter set. We report the median and interquartile range across parameter sets at each grid cell. A response direction is considered robust when at least 17 of the 20 parameter sets agree on the sign of the median response across the five climate models. A strong hotspot is considered robust when at least 17 of the 20 parameter sets classify a grid cell in the same direction (an increase of at least 20\% or a decrease of at least 20\%), with at least four of the five climate models agreeing within each parameter set.

\subsection*{Yield loss attribution}
\label{sec:yield_loss_attribution}
We analyse the 2018 European wheat event with the optimized Agrocosm using automatic differentiation (AD). The reported yields are from the harmonized European Union subnational crop-statistics dataset \cite{ronchetti2023harmonized}. Yield trends are fitted separately to the observed and simulated time series. The 2018 anomalies are calculated as deviations from the corresponding fitted trends following Eqs.~\ref{eq:gridcell_yield_trend} and~\ref{eq:gridcell_yield_anomaly}. The simulated yield anomalies are aggregated to the same administrative units using fixed-2015 wheat area.

For each grid cell, reverse-mode AD calculates the derivatives of end-of-season yield with respect to daily temperature, precipitation, shortwave and longwave radiation along the simulated 2018 crop trajectory. The differentiated period extends from the day after sowing to the day before harvest. Each derivative quantifies the sensitivity of final yield to a single daily weather input. We average the daily derivatives within each month to obtain monthly mean sensitivities. 

To make the four weather sensitivities comparable, we multiply each daily derivative by the standard deviation of its corresponding weather variable over the growing days at each grid cell in 2018. For each day, we add the absolute values of the four scaled sensitivities. We then sum these daily totals over a 14-day window and move the window forward one day at a time. The window with the largest total sensitivity identifies the most weather-sensitive period, and its midpoint represents the timing of peak sensitivity.

Forward counterfactual simulations use weather from 2016, 2017 and 2019 as three separate references. For timing experiments, each simulation replaces all four weather inputs within a single 14-day period with corresponding reference-year values, while retaining 2018 weather on all other days. We advance the period in 14-day steps without overlap and restrict replacement to post-sowing growing days. In driver experiments, we replace one weather variable at a time throughout the post-sowing season. Yield responses are calculated as differences between counterfactual and original 2018 yields. For monthly timing summaries, we average the yield changes from replacement periods whose midpoints fall in the same month, separately for each grid cell and reference year. Regional summaries use crop-area weights over the quarter of wheat area with the most negative reported 2018 anomalies. For counterfactual experiments, regional means are calculated separately for each reference year and summarized by their median and full range.

We compare AD-guided, deficit-based and evenly distributed supplementary irrigation using seasonal water-budget limits of 10--100~mm in 10~mm increments. During the main growing period, the evenly distributed scheme allocates the full budget equally across growing days. The AD-guided scheme uses AD to quantify how the daily crop water deficit responds to additional irrigation and determines the irrigation volume from this sensitivity. The deficit-based scheme instead sets the daily irrigation to the crop water deficit. The latter two schemes make daily decisions using only current crop states and weather, subject to the same seasonal water-budget limit as the evenly distributed scheme. We report actual water use and calculate yield recovery from crop area-weighted yields as the irrigation-induced gain divided by the gap between fitted-trend and 2018 rainfed yields, expressed as a percentage.

\section*{Data availability}

The ISIMIP3 climate forcing and crop management inputs are publicly available for download at \url{https://www.isimip.org/gettingstarted/input-data-bias-adjustment/}.

The daily eddy-covariance flux and meteorological data used for parameter optimization are obtained from FLUXNET products through the FLUXNET Shuttle (\url{https://data.fluxnet.org/}).

The GDHYv1.2+v1.3 benchmark data are downloaded from \url{https://doi.pangaea.de/10.1594/PANGAEA.909132}. The AgMIP GGCMI Phase 3 yield simulations used for comparison are accessible from ISIMIP3 (\url{https://www.isimip.org/outputdata/}).

The harmonized European Union subnational crop statistics used for the 2018
event analysis are available at \url{https://doi.org/10.2905/685949ff-56de-4646-a8df-844b5bb5f835}.

\section*{Code availability}
The source code for Agrocosm is publicly available at \url{https://github.com/yunan-l/Agrocosm.jl}.

\section*{Acknowledgements}

Y.L. acknowledges funding from the program of the China Scholarship Council (no.202303250017). We also acknowledge funding from the ClimTip project. This is ClimTip contribution \#X; the ClimTip project has received funding from the European Union's Horizon Europe research and innovation programme under grant agreement No. 101137601. The authors gratefully acknowledge the Ministry of Research, Science and Culture (MWFK) of Land Brandenburg for supporting this project by providing resources on the high performance computer system at the Potsdam Institute for Climate Impact Research.

\clearpage
\bibliography{sn-bibliography}%

\backmatter
\clearpage

\appendix

\include{supplementary_information}

\end{document}

%% file: supplementary_information.tex
\begin{bibunit}[unsrt]

\section*{\centering Supplementary information}
\renewcommand{\thefigure}{S\arabic{figure}}
\renewcommand{\thetable}{S\arabic{table}}
\renewcommand{\theequation}{S\arabic{equation}}

\renewcommand{\theHfigure}{supp.figure.\arabic{figure}}
\renewcommand{\theHtable}{supp.table.\arabic{table}}
\renewcommand{\theHequation}{supp.equation.\arabic{equation}}
\setcounter{figure}{0}

\addtocontents{toc}{\protect\setcounter{tocdepth}{3}}
\renewcommand{\contentsname}{}
\tableofcontents

\newpage

\section{Model setup}
\label{sec:model_setup}

\subsection{Soil initialization and spin-up}
\label{sec:hwsd_spinup}

The Harmonized World Soil Database (HWSD) provides soil carbon and nitrogen measurements but does not specify their partitioning among Agrocosm's dynamic soil pools \cite{fao2023hwsd_sup}. We convert these measurements to carbon and nitrogen stocks, remap them to Agrocosm's five soil layers, and combine them with static soil properties to initialize the spin-up. A 600-year agricultural spin-up under repeated 1901--1930 weather and fixed 2015 management estimates the fast and slow pool fractions and the vertical routing of organic matter. At the end of each spin-up year, soil organic carbon and total nitrogen in each layer are restored to their HWSD-derived targets, while convergence is assessed across repeated forcing cycles. We then use the spin-up-derived fractions and routing to construct the production initial state from the HWSD stocks.

\subsubsection{HWSD soil inputs}
\label{sec:hwsd_mapping}

Soil organic carbon (SOC), total nitrogen, bulk density, coarse-fragment content and other static soil properties are obtained from HWSD. It resolves seven source layers between 0 and 2~m, whereas Agrocosm uses five prognostic soil layers spanning 0--0.2, 0.2--0.5, 0.5--1.0, 1.0--2.0 and 2.0--3.0~m.

For soil constituent $X\in\{C,N\}$, the stock in HWSD layer $h$ and soil component $k$ is calculated as
\begin{equation}
X^{\mathrm{H}}_{hk}
=
10^{3}q^{X}_{hk}\rho_{hk}d^{\mathrm{H}}_h
\left(1-\frac{CF_{hk}}{100}\right),
\end{equation}
where $q^{C}=OC/100$ and $q^{N}=TN/1000$ are mass fractions, $\rho$ is bulk density in kg~m$^{-3}$, $d^{\mathrm{H}}_h$ is layer thickness in metres and $CF$ is the volumetric percentage of coarse fragments. The resulting stocks are expressed in g~m$^{-2}$.

The reported shares of soil components within each HWSD mapping unit are normalized to sum to one and used to calculate component-weighted stocks. Root-depth classes are used to distinguish structural zeros from unresolved observations: stocks below the represented soil depth are set to zero, whereas missing values within the represented profile remain unresolved. Component weights are not renormalized after applying structural zeros, and a source layer is retained only when at least 99\% of its component share is resolved.

Stocks are transferred from the HWSD layers to the Agrocosm layers by conserving areal stocks under the assumption of uniform stock density within each source layer:
\begin{equation}
\begin{aligned}
X^{\mathrm{mix}}_h
&=
\sum_k w_k X^{\mathrm{H}}_{hk},\\
o_{\ell h}
&=
\left[
\min\left(z^{\mathrm{A}}_{\ell,1},z^{\mathrm{H}}_{h,1}\right)
-
\max\left(z^{\mathrm{A}}_{\ell,0},z^{\mathrm{H}}_{h,0}\right)
\right]_+,\\
X^{\mathrm{A}}_\ell
&=
\sum_h X^{\mathrm{mix}}_h
\frac{o_{\ell h}}{d^{\mathrm{H}}_h},
\end{aligned}
\end{equation}
where $w_k$ is the normalized component share, $o_{\ell h}$ is the vertical overlap between Agrocosm layer $\ell$ and HWSD layer $h$, and $[x]_+=\max(x,0)$. Because HWSD observations end at 2~m, we use the stock density of its deepest layer to initialize Agrocosm's 2--3~m layer. The resulting carbon and nitrogen stocks are extrapolated estimates, not measurements at that depth.

The remapped source pixels are aggregated to the $0.5^{\circ}$ Agrocosm grid using spherical pixel-area weights. A grid-cell layer is retained only when valid-area coverage is at least 99\% for both carbon and nitrogen. Coverage limitations, structural-depth assumptions and deep-soil extrapolation are retained as quality-control flags.

\subsubsection{Construction of initial soil states}
\label{sec:initial_soil_state}

The remapped SOC and total-nitrogen stocks define the layer-specific targets $C^{*}_{\ell c}$ and $N^{*}_{\ell c}$ for grid cell $c$. Before spin-up-derived allocations are available, SOC is divided between fast and slow pools using a 40:60 split:
\begin{equation}
C^{\mathrm{fast}}_{\ell c,0}
=
0.4C^{*}_{\ell c},
\qquad
C^{\mathrm{slow}}_{\ell c,0}
=
0.6C^{*}_{\ell c}.
\end{equation}
Organic nitrogen is initially divided using the same ratio. Nitrate and ammonium are each initialized at 1\% of the slow organic-nitrogen pool, and the organic-nitrogen total is adjusted so that the sum of the organic and mineral pools equals the HWSD total-nitrogen target. Initial litter carbon and nitrogen are set to zero. Before spin-up-derived routing is available, 55\% of vertically routed carbon is assigned to the uppermost layer and the remaining 45\% is divided equally among the four deeper layers. These initial fractions are used only to start the spin-up and are subsequently replaced by spin-up-derived allocations.

Initial liquid-water storage is set to field capacity:
\begin{equation}
W_{\ell c,0}
=
\theta^{\mathrm{FC}}_{\ell c}d^{\mathrm{A}}_\ell,
\end{equation}
where $\theta^{\mathrm{FC}}_{\ell c}$ is volumetric field capacity and $d^{\mathrm{A}}_\ell$ is layer thickness expressed in millimetres. Field capacity is estimated from sand and clay fractions and SOC-derived organic matter using the Saxton--Rawls pedotransfer relationships \cite{saxton2006soil}. Static soil properties are supplied separately to the hydraulic and thermal process calculations.

\subsubsection{Spin-up and convergence criteria}
\label{sec:agricultural_spin}

We conduct a 600-year target-constrained agricultural spin-up using the complete crop, carbon, nitrogen, water and thermal process calculations. Management is fixed at its 2015 values \cite{frieler2024scenario}, while daily climate from 1901--1930 is repeated over successive 30-year cycles using a 365-day calendar. Monthly 2015 nitrate and ammonium deposition is repeated annually, interpolated to daily rates and applied to the mineral-nitrogen pools in the uppermost soil layer.

At the end of each simulated year, the soil pools in every layer are proportionally rescaled to the HWSD-derived SOC and total-nitrogen targets. For target stock $S^\star_{\ell c}$ and constituent pool $j$, the stock after proportional rescaling is
\begin{equation}
S^{+}_{j,\ell c,y}
=
\frac{S^{*}_{\ell c}}
{\displaystyle\sum_{j'\in\mathcal{P}_{S}}
\widetilde{S}_{j',\ell c,y}}
\widetilde{S}_{j,\ell c,y},
\end{equation}
where tildes denote stocks immediately before correction and the superscript $+$ denotes stocks after rescaling, not amounts added to the original stocks, 
$\mathcal{P}_{C}=\{\mathrm{fast},\mathrm{slow}\}$ and
$\mathcal{P}_{N}=\{\mathrm{fast},\mathrm{slow},\mathrm{NO_3},\mathrm{NH_4}\}$.
This correction preserves the simulated relative allocation among pools while preventing total soil stocks from drifting away from the HWSD targets. If the pre-correction stock is numerically zero, the provisional allocation is restored. Litter carbon and nitrogen remain unconstrained and evolve entirely through the simulated processes. The correction is applied annually throughout spin-up rather than as a single adjustment after year 600.

Convergence is evaluated by comparing states at the same phase of successive 30-year climate cycles. The diagnostics comprise total carbon and nitrogen, including litter; the fast-pool fractions of organic carbon and nitrogen; and changes in the annual carbon and nitrogen target corrections. A grid cell is considered stable when relative changes in total stocks and normalized corrections are no greater than 0.01 and absolute changes in fast-pool fractions are no greater than 0.01. These conditions must hold for three consecutive same-phase comparisons. The fraction of cells satisfying all criteria is retained as a spin-up quality-control metric.

We combine the fast-pool fractions and vertical-routing fractions derived during spin-up with the HWSD-derived stocks and static soil properties to initialize production simulations. Unlike during spin-up, no annual correction to the HWSD targets is applied during production.

\subsection{Spin-up of four crops}
\label{sec:spinup_results}

\begin{table}[ht]
\centering
\caption{\textbf{Per-cell convergence of the 600-year target-constrained spin-up for all four crops under both rainfed and irrigated conditions} A grid cell is classified as converged when changes in total carbon and nitrogen stocks, fast-pool fractions and normalized target corrections remain within their prescribed tolerances for three consecutive comparisons with the corresponding years of the previous 30-year forcing cycle.}
\label{tab:warmup_convergence}
\begin{tabular}{llrrr}
\hline
Crop & Water system & Active cells & Converged cells & Rate (\%) \\
\hline
Wheat   & Rainfed   & 32\,967 & 32\,940 &  99.92 \\
Wheat   & Irrigated &  6\,579 &  6\,579 & 100.00 \\
Maize   & Rainfed   & 34\,520 & 34\,500 &  99.94 \\
Maize   & Irrigated &  7\,560 &  7\,560 & 100.00 \\
Rice    & Rainfed   & 15\,698 & 15\,686 &  99.92 \\
Rice    & Irrigated &  7\,589 &  7\,588 &  99.99 \\
Soybean & Rainfed   & 25\,233 & 25\,214 &  99.92 \\
Soybean & Irrigated &  4\,312 &  4\,312 & 100.00 \\
\hline
\end{tabular}
\end{table}

\clearpage

\section{FLUXNET observations}
\label{sec:fluxnet_data}

We use daily flux and meteorological products from the FLUXNET products \cite{pastorello2020fluxnet2015_sup}. The observational constraints are gross primary production (GPP), ecosystem respiration (RECO)
and total evapotranspiration (ET). Daily GPP (\text{GPP\_NT\_VUT\_REF}) and RECO (\text{RECO\_NT\_VUT\_REF}) are taken from the night-time flux partitioning products \cite{reichstein2005separation}. Daily ET is derived from the gap-filled latent heat flux (\text{LE\_F\_MDS}) and air temperature:
\begin{equation}
ET_t =
\frac{LE_t \times 86400}
{2.501\times10^{6}-2361T_{\mathrm{A},t}},
\end{equation}
where $LE_t$ is in W m$^{-2}$, $T_{\mathrm{A},t}$ is air temperature in degrees Celsius and $ET_t$ is in mm d$^{-1}$.

Each candidate season is checked for complete daily coverage of the five forcing variables used by Agrocosm: air temperature (\text{TA\_F}), precipitation (\text{P\_F}), incoming longwave radiation (\text{LW\_IN\_F}), incoming shortwave radiation (\text{SW\_IN\_F}) and gap-filled atmospheric CO$_2$ (\text{CO2\_F\_MDS}). Short gaps in CO$_2$ are replaced using the same no-leap calendar day from the preceding, following or nearest available year.

\begin{figure}[H]
\centering
\includegraphics[width=\textwidth]{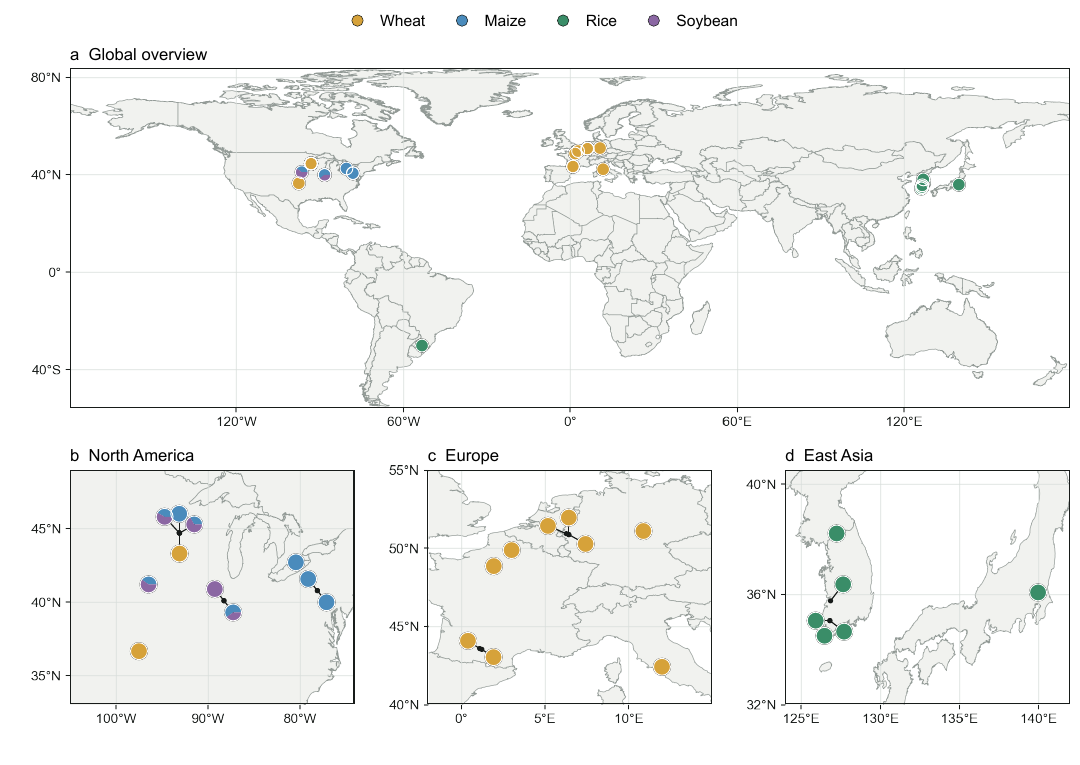}
\caption{\textbf{Global and regional distribution of flux-tower sites used for parameter optimization.}
\textbf{a}, Global overview of the 27 physical sites. 
\textbf{b--d}, Enlarged views of North America, Europe and East Asia, respectively. 
The dataset comprises 100 growing-season records from 27 physical sites, including 36 wheat, 22 maize, 26 rice and 16 soybean records. Colours denote the target crops, with segmented markers indicating sites contributing observations for multiple crops.}
\label{fig:flux_site_distribution}
\end{figure}

Crop identity and sowing dates are resolved from station metadata and crop-rotation records. GPP, RECO and ET are compared only within the sowing-to-harvest window. Station coordinates are matched to the nearest Agrocosm grid cell. The final calibration set contains 100 growing season records from 27 physical sites: 36 wheat records from 11 sites, 22 maize records from 8 sites, 26 rice records from 7 sites and 16 soybean records from 5 sites. Site coordinates, crop coverage, season counts and available metadata are provided in Table~\ref{tab:fluxnet_site_metadata}, and the spatial distribution of sites is shown in Fig.~\ref{fig:flux_site_distribution}.

\begingroup
\small
\setlength{\tabcolsep}{5pt}
\begin{longtable}{p{1.4cm}p{2.8cm}rrcp{2.6cm}}
\caption{Site identifiers, coordinates and target-crop coverage for the 27 physical flux sites retained in the final calibration set. The set comprises 100 growing-season records; numbers in parentheses denote the records contributed by each site and crop.}
\label{tab:fluxnet_site_metadata}\\
\hline
Site ID & Site name & Latitude & Longitude & Target crop (n) \\
\hline
\endfirsthead
\hline
Site ID & Site name & Latitude & Longitude & Target crop (n) \\
\hline
\endhead
\hline
\endfoot
\hline
\endlastfoot
BR-SM1 & Rice Paddy & -30.28 & -53.15 & rice (1) \\
CA-TPA & Agricultural Site & 42.69 & -80.35 & maize (1) \\
DE-Geb & Gebesee & 51.10 & 10.91 & wheat (4) \\
DE-RuM & Merzenhausen & 50.93 & 6.30 & wheat (3) \\
DE-RuS & Selhausen Juelich & 50.87 & 6.45 & wheat (6) \\
DE-Seh & Selhausen & 50.87 & 6.45 & wheat (2) \\
FR-Aur & Aurade & 43.55 & 1.11 & wheat (6) \\
FR-EM2 & Estrees-Mons A28 & 49.87 & 3.02 & wheat (2) \\
FR-Gri & Grignon & 48.84 & 1.95 & wheat (3) \\
FR-Lam & Lamasquere & 43.50 & 1.24 & wheat (2) \\
IT-CA2 & Castel d'Asso2 & 42.38 & 12.03 & wheat (2) \\
JP-Mse & Mase rice paddy field & 36.05 & 140.03 & rice (8) \\
KR-CRK & Cheorwon Rice paddy & 38.20 & 127.25 & rice (7) \\
KR-GmR & Gimje Rice Paddy & 35.75 & 126.86 & rice (3) \\
KR-HPK & Haenam Paddy-field & 34.48 & 126.48 & rice (3) \\
KR-Nj1 & Naju Rice Paddy 1 & 35.03 & 126.82 & rice (2) \\
KR-Nj2 & Naju Rice Paddy 2 & 35.03 & 126.82 & rice (2) \\
US-ARM & Great Plains site & 36.61 & -97.49 & wheat (5) \\
US-Ne3 & Maize--soybean site & 41.18 & -96.44 & maize (2), soybean (3) \\
US-Ro1 & Rosemount--G21 & 44.71 & -93.09 & maize (5), soybean (7) \\
US-Ro3 & Rosemount--G19 & 44.72 & -93.09 & maize (1) \\
US-Ro5 & Rosemount I18--South & 44.69 & -93.06 & maize (1), soybean (2) \\
US-Ro6 & Rosemount I18--North & 44.69 & -93.06 & wheat (1) \\
US-UC1 & LTAR UCB EC1 & 40.75 & -78.01 & maize (4) \\
US-UC2 & LTAR UCB EC2 & 40.76 & -78.00 & maize (5) \\
US-UiE & Sorghum--Soy site & 40.06 & -88.20 & soybean (2) \\
US-UiG & Maize--Soy site & 40.06 & -88.20 & maize (3), soybean (2) \\
\end{longtable}
\endgroup

\clearpage
\section{Evaluation metrics}\label{sec:evaluation_metrics}

\subsection*{Station-flux metrics} 

For each growing season $s$ and flux $q$, Pearson correlation is calculated across the valid observation days $\mathcal{T}_{s,q}$:
\begin{equation}
r_{s,q}=
\frac{\sum_{t\in\mathcal{T}_{s,q}}(y^{\mathrm{model}}_{s,q,t}-\overline{y}^{\mathrm{model}}_{s,q})(y^{\mathrm{obs}}_{s,q,t}-\overline{y}^{\mathrm{obs}}_{s,q})}
{\sqrt{\sum_{t\in\mathcal{T}_{s,q}}(y^{\mathrm{model}}_{s,q,t}-\overline{y}^{\mathrm{model}}_{s,q})^2}\sqrt{\sum_{t\in\mathcal{T}_{s,q}}(y^{\mathrm{obs}}_{s,q,t}-\overline{y}^{\mathrm{obs}}_{s,q})^2}}.
\end{equation}
The normalized root-mean-square error uses the observation scale $\sigma_{s,q}$ defined in section~\ref{sec:parameter_selection}:
\begin{equation}
\operatorname{nRMSE}_{s,q}
=
\frac{1}{\sigma_{s,q}}
\sqrt{
\frac{1}{N_{s,q}}
\sum_{t\in\mathcal{T}_{s,q}}
\left(y^{\mathrm{model}}_{s,q,t}-y^{\mathrm{obs}}_{s,q,t}\right)^2
}.
\end{equation}

\subsection*{Global yield metrics}

For parameterization $s$, crop $c$, grid cell $g$ and year $t$, the combined grid-cell total yield $Y^{s}_{g,c,t}$ is calculated as the crop area-weighted mean of rainfed and irrigated yields:
\begin{equation}
Y^{s}_{g,c,t}
=
\frac{
p^{\mathrm{rf}}_{g,c}Y^{s,\mathrm{rf}}_{g,c,t}
+
p^{\mathrm{irr}}_{g,c}Y^{s,\mathrm{irr}}_{g,c,t}
}{
p^{\mathrm{rf}}_{g,c}+p^{\mathrm{irr}}_{g,c}
}.
\end{equation}

Using the detrended anomalies defined in Methods (Eqs.~\ref{eq:gridcell_yield_trend} and~\ref{eq:gridcell_yield_anomaly}), Pearson temporal correlation over the valid years $\mathcal{T}_{g,c}$ is
\begin{equation}
r^{s}_{g,c}
=
\frac{
\sum_{t\in\mathcal{T}_{g,c}}
\widetilde{Y}^{s}_{g,c,t}\widetilde{Y}^{\mathrm{GDHY}}_{g,c,t}
}{
\sqrt{\sum_{t\in\mathcal{T}_{g,c}}(\widetilde{Y}^{s}_{g,c,t})^2}
\sqrt{\sum_{t\in\mathcal{T}_{g,c}}(\widetilde{Y}^{\mathrm{GDHY}}_{g,c,t})^2}
}.
\end{equation}
The root-mean-square error (RMSE) is
\begin{equation}
\operatorname{RMSE}^{s}_{g,c}
=
\sqrt{
\frac{1}{|\mathcal{T}_{g,c}|}
\sum_{t\in\mathcal{T}_{g,c}}
\left(
\widetilde{Y}^{s}_{g,c,t}
-
\widetilde{Y}^{\mathrm{GDHY}}_{g,c,t}
\right)^2
},
\end{equation}
and normalized RMSE is
\begin{equation}
\operatorname{nRMSE}^{s}_{g,c}
=
\frac{
\sqrt{|\mathcal{T}_{g,c}|^{-1}
\sum_{t\in\mathcal{T}_{g,c}}
(\widetilde{Y}^{s}_{g,c,t}-\widetilde{Y}^{\mathrm{GDHY}}_{g,c,t})^2}
}{
\operatorname{SD}(\widetilde{Y}^{\mathrm{GDHY}}_{g,c})
},
\end{equation}
where $\operatorname{SD}(\widetilde{Y}^{\mathrm{GDHY}}_{g,c})$ is the standard deviation of detrended GDHY yield anomalies across the evaluation years $\mathcal{T}_{g,c}$ for crop $c$ in grid cell $g$.

Low-yield years, denoted by $\mathcal{L}_{g,c}$, are the years in the lowest 20\% of detrended GDHY yield anomalies within $\mathcal{T}_{g,c}$. The corresponding mean absolute error (MAE) is
\begin{equation}
\operatorname{MAE}^{s,\mathrm{low}}_{g,c}
=
\frac{1}{|\mathcal{L}_{g,c}|}
\sum_{t\in\mathcal{L}_{g,c}}
\left|
\widetilde{Y}^{s}_{g,c,t}
-
\widetilde{Y}^{\mathrm{GDHY}}_{g,c,t}
\right|,
\end{equation}
and the normalized MAE in low-yield years is
\begin{equation}
\operatorname{nMAE}^{s,\mathrm{low}}_{g,c}
=
\frac{
\operatorname{MAE}^{s,\mathrm{low}}_{g,c}
}{
\operatorname{SD}_{t\in\mathcal{T}_{g,c}}
\left(\widetilde{Y}^{\mathrm{GDHY}}_{g,c,t}\right)
}.
\end{equation}

\clearpage
\section{Parameter estimation}
\label{sec:parameter_selection}

\subsection{Parameter selection}

We first define a candidate set of crop and soil parameters covering canopy development, radiation interception, photosynthesis, respiration, canopy interception, nitrogen allocation, soil-carbon turnover, evaporation and transpiration. Discrete management decisions, including sowing, harvest and crop-failure events, are kept fixed during parameter calibration and are resolved before the differentiable daily transition.

Parameter sensitivity is evaluated from the normalized Jacobian of the differentiable Agrocosm trajectory. For each growing season $s$, diagnostic flux $q$ and valid observation day $t$, we define the observation scale as
\begin{equation}
\sigma_{s,q} =
\max\left[
\operatorname{sd}\left(y^{\mathrm{obs}}_{s,q,t}\right),
\frac{1}{N_{s,q}}\sum_{t\in\mathcal{T}_{s,q}}
\left|y^{\mathrm{obs}}_{s,q,t}\right|,
10^{-6}
\right],
\end{equation}
where $\mathcal{T}_{s,q}$ denotes the days with valid observations and $N_{s,q}=|\mathcal{T}_{s,q}|$. Each parameter is expressed in normalized bounded coordinates,
\begin{equation}
z_i=\frac{\theta_i-\ell_i}{u_i-\ell_i},
\qquad
\theta_i=\ell_i+z_i(u_i-\ell_i),
\qquad 0\leq z_i\leq 1,
\end{equation}
where $\ell_i$ and $u_i$ are the lower and upper bounds. The normalized sensitivity matrix is calculated as
\begin{equation}
J^{(s)}_{(q,t),i}
=
\frac{\partial}{\partial z_i}
\left[
\frac{y^{\mathrm{model}}_{s,q,t}
\left(\boldsymbol{\theta}(\boldsymbol{z})\right)}
{\sigma_{s,q}}
\right].
\end{equation}

The candidate parameters are evaluated using the norms and singular-value structure of the Jacobian columns, pairwise parameter correlations, multi-start optimization and leave-one-parameter-out experiments. Parameters are retained when they produce finite and reproducible gradients and contribute independent information across seasons and diagnostic fluxes.

The final calibration contains 12 crop parameters (Table~\ref{tab:cft_parameters}) and four soil parameters (Table~\ref{tab:soil_parameters}) for each crop. The two parameter vectors are fitted independently for wheat (winter and spring wheat), maize, rice and soybean.

\begin{table*}[!htbp]
\renewcommand{\arraystretch}{1.5}
\centering
\newcommand{\boundcell}[2]{%
\parbox[c]{0.16\linewidth}{\centering #1\par #2}}
\caption{Crop physiological parameters and calibration bounds. The bounds define optimization search ranges rather than universal limits. W, wheat; R, rice; M, maize; S, soybean. Bounds for $\text{basetemp\_low}$ are in $^\circ$C.}
\label{tab:cft_parameters}
\resizebox{0.95\linewidth}{!}{%
\begin{tabular}{p{0.15\linewidth}p{0.41\linewidth}cc}
\hline
Parameter & Description & Lower bound & Upper bound \\
\hline
\text{laimax} & Maximum potential leaf-area index and canopy development &
\raisebox{-12pt}{\shortstack[c]{3.5 (W)\quad 5.0 (R)\\4.0 (M)\quad 5.0 (S)}} &
\raisebox{-12pt}{\shortstack[c]{9.0 (W)\quad 9.0 (R)\\9.0 (M)\quad 9.0 (S)}} \\
\text{sla} & Specific leaf area, converting leaf carbon to leaf area & 0.015 & 0.070 \\
\text{alphaa} & Coefficient scaling absorbed photosynthetically active radiation entering photosynthesis & 0.40 & 1.00 \\
\text{lightextcoeff} & Canopy light-extinction coefficient controlling the fraction of absorbed radiation & 0.30 & 1.10 \\
\text{gmin} & Minimum canopy conductance and lower bound on crop gas exchange & 0.30 & 2.00 \\
\text{b} & Dark-respiration coefficient relative to photosynthetic capacity & 0.010 & 0.080 \\
\text{respcoeff} & Maintenance respiration coefficient for below-ground and storage pools & 0.10 & 1.50 \\
\text{intc} & Canopy-interception storage coefficient per unit leaf area & 0.005 & 0.100 \\
\text{knstore} & Nitrogen-storage allocation coefficient regulating reserve formation and nitrogen stress & 0.00 & 0.30 \\
\text{albedo\_leaf} & Leaf shortwave albedo controlling the absorbed radiation available for photosynthesis & 0.08 & 0.30 \\
\text{basetemp\_low} & Base temperature for heat-unit accumulation during crop development &
\raisebox{-12pt}{\shortstack{$-2$ (W) 5 (R)\\3 (M) 4 (S)}} &
\raisebox{-12pt}{\shortstack{5 (W) 12 (R)\\10 (M) 10 (S)}} \\
\text{fphusen} & Fraction of accumulated potential heat units at the onset of senescence &
\raisebox{-12pt}{\shortstack[c]{0.60 (W)\quad 0.65 (R)\\0.60 (M)\quad 0.60 (S)}} &
\raisebox{-12pt}{\shortstack[c]{0.90 (W)\quad 0.92 (R)\\0.90 (M)\quad 0.90 (S)}} \\
\hline
\end{tabular}
}
\end{table*}

\begin{table*}[!htbp]
\renewcommand{\arraystretch}{1.5}
\centering
\caption{Soil parameters and calibration bounds. The bounds define optimization search ranges rather than universal limits.}
\label{tab:soil_parameters}
\resizebox{\textwidth}{!}{%
\begin{tabular}{p{0.3\linewidth}p{0.38\linewidth}cc}
\hline
Parameter & Description & Lower bound & Upper bound \\
\hline
\text{k\_soil10\_fast} & Daily decomposition rate of the fast soil carbon pool at $10\,^{\circ}\mathrm{C}$ & $2\times10^{-5}$ & $3\times10^{-4}$ \\
\text{PRIESTLEY\_TAYLOR} & Coefficient scaling equilibrium and potential evaporation & 0.8 & 1.8 \\
\text{soildepth\_evap} & Effective soil depth contributing to soil evaporation & $100$ & $500$ \\
\text{soil\_infil} & Infiltration control governing the partitioning of rainfall into soil infiltration and runoff & 1.0 & 3.0 \\
\hline
\end{tabular}
}
\end{table*}

\subsection{Two-stage calibration}
\label{sec:two_stage_calibration}

The simultaneous optimization of all crop and soil parameters produced less stable fits and poorer performance (not shown here). We therefore use two-stage optimization to reduce the effective search dimension and limit compensation between crop and soil parameters. For each growing season and diagnostic flux, the normalized mean-squared error is calculated as
\begin{equation}
\ell_{s,q}(\boldsymbol{\theta})
=
\frac{1}{N_{s,q}}
\sum_{t\in\mathcal{T}_{s,q}}
\left[
\frac{
y^{\mathrm{model}}_{s,q,t}(\boldsymbol{\theta})
-
y^{\mathrm{obs}}_{s,q,t}
}{
\sigma_{s,q}
}
\right]^2 .
\end{equation}
In the first calibration stage, the 12 crop parameters are optimized independently for each crop $c$:
\begin{equation}
\mathcal{L}_{\mathrm{CFT}}^{(c)}
(\boldsymbol{\theta}_c)
=
\sum_{s\in S_c}
w^{\mathrm{CFT}}_s
\left[
\ell_{s,\mathrm{GPP}}
+
\ell_{s,\mathrm{RECO}}
+
\ell_{s,\mathrm{ET}}
\right],
\label{eq:cft_objective}
\end{equation}
where $S_c$ is the set of site--seasons for crop $c$. The weighting
\[
w^{\mathrm{CFT}}_s
=
\frac{1}{
N_{\mathrm{site},c}
n_{\mathrm{season}(s)}
}
\]
gives equal total weight to each physical site within a crop and prevents sites with more years from dominating the objective.

After the crop-specific parameters $\widehat{\boldsymbol{\theta}}_c$ are obtained, they are held fixed in the second stage. The four shared soil parameters $\boldsymbol{\phi}$ are then optimized jointly across all crops:
\begin{equation}
\mathcal{L}_{\mathrm{soil}}(\boldsymbol{\phi})
=
\sum_{c}
\sum_{s\in S_c}
w^{\mathrm{soil}}_s
\left[
\ell_{s,\mathrm{RECO}}
\left(\widehat{\boldsymbol{\theta}}_c,\boldsymbol{\phi}\right)
+
\ell_{s,\mathrm{ET}}
\left(\widehat{\boldsymbol{\theta}}_c,\boldsymbol{\phi}\right)
\right].
\label{eq:soil_objective}
\end{equation}
The soil-stage weights assign equal total weight to each crop, each physical site within a crop and each season within a site:
\[
w^{\mathrm{soil}}_s
=
\frac{1}{
N_{\mathrm{crop}}
N_{\mathrm{site},c}
n_{\mathrm{season}(c,s)}
}.
\]
GPP is not included in the soil-stage objective and is retained as an independent diagnostic for evaluating the effect of the shared soil parameters.

Optimization is performed in the normalized coordinates $\boldsymbol{z}$. For an objective gradient $\boldsymbol{g}_k=
\nabla_{\boldsymbol{z}}\mathcal{L}_k$, Adam \cite{kingma2015adam_sup} updates are calculated as
\begin{align}
\boldsymbol{m}_k &=
\beta_1\boldsymbol{m}_{k-1}
+
(1-\beta_1)\boldsymbol{g}_k,\\
\boldsymbol{v}_k &=
\beta_2\boldsymbol{v}_{k-1}
+
(1-\beta_2)\boldsymbol{g}_k^2,
\end{align}
with bias-corrected moments
\[
\widehat{\boldsymbol{m}}_k
=
\frac{\boldsymbol{m}_k}{1-\beta_1^k},
\qquad
\widehat{\boldsymbol{v}}_k
=
\frac{\boldsymbol{v}_k}{1-\beta_2^k}.
\]
The normalized parameters are updated according to
\begin{equation}
\boldsymbol{z}_{k+1}
=
\operatorname{clip}
\left[
\boldsymbol{z}_k
-
\eta
\frac{\widehat{\boldsymbol{m}}_k}
{\sqrt{\widehat{\boldsymbol{v}}_k}+\epsilon},
0,1
\right].
\end{equation}
The Adam parameters are $\beta_1=0.9$, $\beta_2=0.999$ and $\epsilon=10^{-8}$. The full-data calibration used a learning rate of $\eta=0.01$, one growing season per batch, 100 epochs for the crop stage and 20 epochs for the soil stage. Reverse-mode checkpointing is applied in 30-day blocks to limit the memory required for trajectory differentiation.

\subsection{Grouped-site cross-validation}
\label{sec:grouped_site_cross_validation}

We use grouped-site fivefold cross-validation (Table~\ref{tab:site_folds}). All seasons and crops from a given physical site are assigned to the same fold, preventing site-specific soil and climate information from being shared between training and validation data.

\begin{table}[!h]
\centering
\caption{Held-out physical-site groups used for fivefold cross-validation. Crop columns report the number of held-out site--seasons.}
\label{tab:site_folds}
\begin{tabular}{lrrrrrr}
\hline
Fold & Sites & Seasons & Maize & Rice & Soybean & Wheat \\
\hline
1 & 4 & 21 & 5 & 3 & 7 & 6 \\
2 & 5 & 20 & 3 & 8 & 2 & 7 \\
3 & 5 & 21 & 5 & 7 & 2 & 7 \\
4 & 6 & 19 & 5 & 4 & 2 & 8 \\
5 & 7 & 19 & 4 & 4 & 3 & 8 \\
\hline
Total & 27 & 100 & 22 & 26 & 16 & 36 \\
\hline
\end{tabular}
\end{table}

For each fold, both calibration stages are refitted from their default values using only the training sites. The resulting crop and soil parameters are held fixed and evaluated on the corresponding held-out sites without further parameter updates. These fold-specific parameter products are only used to quantify cross-site generalization. 

\begin{figure}[H]
\centering
\includegraphics[width=\textwidth]{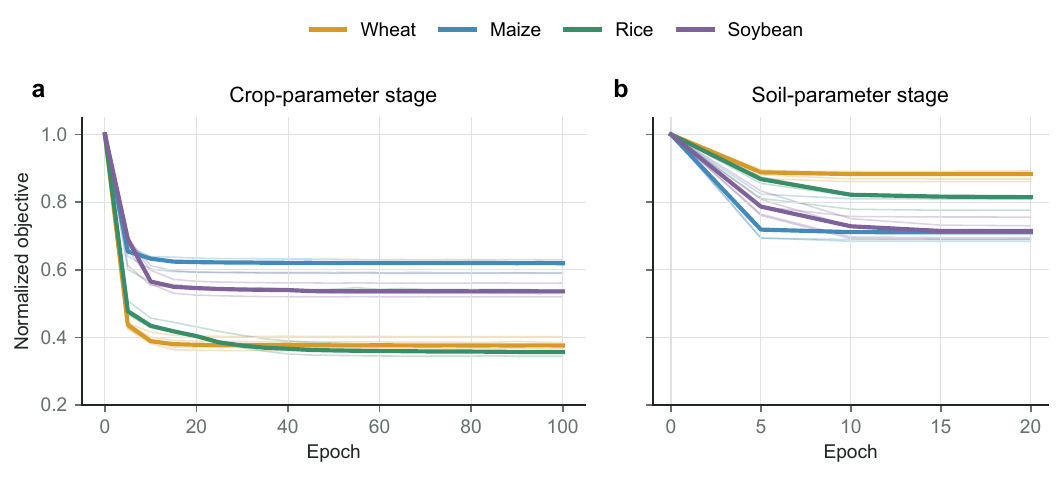}
\caption{\textbf{Optimization convergence across five folds.}
\textbf{a,b}, Weighted training losses for the crop- and soil-parameter stages, respectively. The crop-stage loss combines normalized squared errors in GPP, RECO and ET (Eq.~\eqref{eq:cft_objective}). The soil-stage loss combines those in RECO and ET (Eq.~\eqref{eq:soil_objective}). Each loss is normalized relative to its value at epoch 0 to show its relative reduction. Thin lines show individual fits for each crop and fold. Thick lines show the median across five folds for each crop. Losses are evaluated every five epochs.}
\label{fig:cv5_optimization_diagnostics}
\end{figure}

In all 20 fits in each stage (four crops across five folds), the final weighted training objective is lower than its value at epoch 0 (Fig.~\ref{fig:cv5_optimization_diagnostics}). These are the crop-stage objective $\mathcal{L}_{\mathrm{CFT}}^{(c)}$ (Eq.~\eqref{eq:cft_objective}) and the soil-stage objective $\mathcal{L}_{\mathrm{soil}}$ (Eq.~\eqref{eq:soil_objective}) defined above. Across crops, the mean loss reduction ranges from 38.4\% to 62.2\% for the crop parameters and from 12.3\% to 28.3\% for the soil parameters. The final reported losses remain within 0.05\% of the best evaluated epoch on average, indicating consistent convergence across folds and crops.

\begin{figure}[H]
\centering
\includegraphics[width=0.92\textwidth]{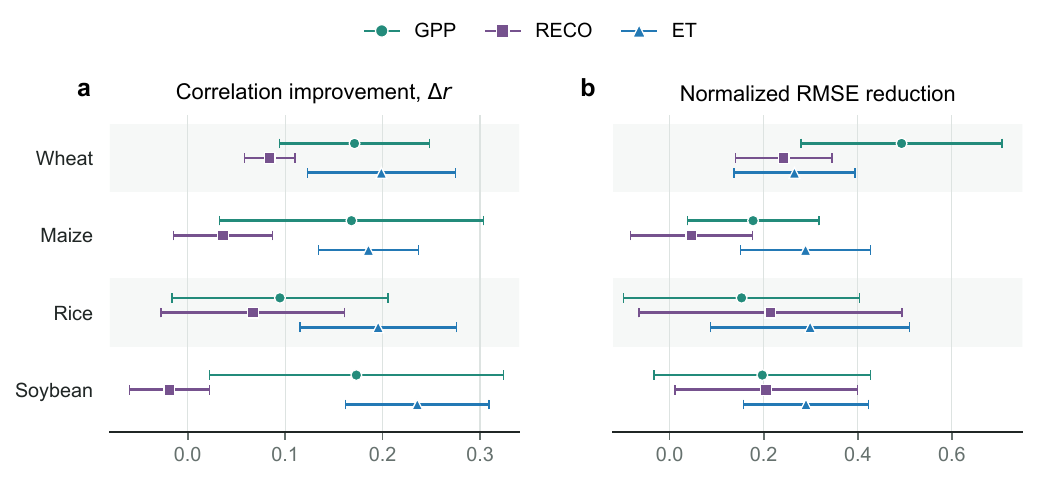}
\caption{\textbf{Cross-site generalization.}
Points show the site-equal mean improvement from the default to optimized parameterization, and horizontal lines show 95\% confidence intervals from 20,000 physical-site cluster bootstrap replicates. Correlation improvement is optimized value minus default value, whereas nRMSE reduction is default minus optimized. Positive values therefore indicate improvement in both panels.}
\label{fig:cv5_crop_generalization}
\end{figure}

For each metric, we first average its season-level values separately for each crop at a physical site. At sites with multiple crops, we then average these crop-specific metric values to obtain one value per site. Overall performance is the unweighted mean of the 27 site-level values (Table~\ref{tab:cv5_primary_results}), so each physical site contributes equally regardless of its number of crops or growing seasons. The results show strong GPP and ET gains across all four crops, while the RECO response is less uniform (Fig.~\ref{fig:cv5_crop_generalization}). Wheat improves consistently across all fluxes. Maize RECO shows the smallest nRMSE reduction (7.3\%). Rice RECO correlation increases only modestly ($\Delta r=0.067$). Separating the two calibration stages shows that the crop parameters account for most GPP and RECO improvements, while the soil parameters provide a comparable contribution to ET correlation (Table~\ref{tab:cv5_stage_contributions}). The soil-parameter calibration increases ET correlation by a further 0.09 with a positive increment at all 27 physical sites.

\begin{table*}[t]
\centering
\small
\caption{Grouped-site cross-validation performance across 27 held-out physical sites. Values are site-equal means.}
\label{tab:cv5_primary_results}
\resizebox{\textwidth}{!}{%
\begin{tabular}{lrrrrrrrr}
\hline
Flux & \multicolumn{4}{c}{Correlation} & \multicolumn{4}{c}{Normalized RMSE} \\
Flux & default & optimized & $\Delta r$ & Improved sites & default & optimized & Reduction & Improved sites \\
\hline
GPP  & 0.649 & 0.807 & 0.157 & 25/27 & 0.897 & 0.584 & 34.9\% & 25/27 \\
RECO & 0.649 & 0.708 & 0.059 & 24/27 & 0.748 & 0.562 & 24.8\% & 23/27 \\
ET   & 0.537 & 0.737 & 0.200 & 27/27 & 0.766 & 0.489 & 36.2\% & 27/27 \\
\hline
\end{tabular}
}
\end{table*}

\begin{table}[t]
\centering
\caption{Contributions of the crop- and soil-parameter stages to overall held-out improvement. Positive values denote improvement.}
\label{tab:cv5_stage_contributions}
\begin{tabular}{llrrr}
\hline
Flux & Metric & Crop parameters & Soil parameters & Total \\
\hline
GPP  & $\Delta r$       & 0.149 & 0.008 & 0.157 \\
GPP  & nRMSE reduction  & 0.300 & 0.013 & 0.313 \\
RECO & $\Delta r$       & 0.055 & 0.004 & 0.059 \\
RECO & nRMSE reduction  & 0.164 & 0.021 & 0.186 \\
ET   & $\Delta r$       & 0.112 & 0.088 & 0.200 \\
ET   & nRMSE reduction  & 0.202 & 0.076 & 0.278 \\
\hline
\end{tabular}
\end{table}

\clearpage

\subsection{Full-data parameter optimization}
\label{sec:full_data_parameterization}

After cross-validation, we re-train separate parameter sets for winter wheat, spring wheat, maize, rice and soybean using all 100 growing seasons. We then use these parameter sets in the global simulations. The final parameter values are reported in Tables~\ref{tab:final_crop_parameterization} and~\ref{tab:final_soil_parameterization}.

\begin{longtable}{p{2.1cm}p{3.0cm}r@{\hspace{0.8cm}}rp{1.8cm}}
\caption{Final full-data crop parameterization used in the global simulations. Parameter values are shown to three significant digits. Simulations use the full-precision values.}\label{tab:final_crop_parameterization}\\
\hline
Parameter product & Parameter & Default & Final & Status \\
\hline
\endfirsthead
\multicolumn{5}{l}{\tablename\ \thetable\ continued}\\
\hline
Parameter product & Parameter & Default & Final & Status \\
\hline
\endhead
Winter wheat & \texttt{laimax} & 7 & 4.00 & Fitted \\
 & \texttt{sla} & 0.0365 & 0.07 & Fitted \\
 & \texttt{alphaa} & 1 & 0.419 & Fitted \\
 & \texttt{lightextcoeff} & 0.5 & 1.1 & Fitted \\
 & \texttt{gmin} & 1 & 0.3 & Fitted \\
 & \texttt{b} & 0.031 & 0.08 & Fitted \\
 & \texttt{respcoeff} & 0.8 & 0.8 & Fixed \\
 & \texttt{intc} & 0.02 & 0.02 & Fixed \\
 & \texttt{knstore} & 0.1 & 0.1 & Fixed \\
 & \texttt{albedo\_leaf} & 0.18 & 0.18 & Fixed \\
 & \texttt{basetemp\_low} & 0 & 1.37 & Fitted \\
 & \texttt{fphusen} & 0.7 & 0.607 & Fitted \\
\addlinespace[2pt]
Spring wheat & \texttt{laimax} & 7 & 3.70 & Fitted \\
 & \texttt{sla} & 0.0365 & 0.0365 & Fixed \\
 & \texttt{alphaa} & 1 & 1 & Fixed \\
 & \texttt{lightextcoeff} & 0.5 & 0.3 & Fitted \\
 & \texttt{gmin} & 1 & 1 & Fixed \\
 & \texttt{b} & 0.031 & 0.0573 & Fitted \\
 & \texttt{respcoeff} & 0.8 & 0.8 & Fixed \\
 & \texttt{intc} & 0.02 & 0.02 & Fixed \\
 & \texttt{knstore} & 0.1 & 0.1 & Fixed \\
 & \texttt{albedo\_leaf} & 0.18 & 0.18 & Fixed \\
 & \texttt{basetemp\_low} & 0 & 0 & Fixed \\
 & \texttt{fphusen} & 0.7 & 0.650 & Fitted \\
\addlinespace[2pt]
Maize & \texttt{laimax} & 5 & 7.85 & Fitted \\
 & \texttt{sla} & 0.0431 & 0.0431 & Fixed \\
 & \texttt{alphaa} & 1 & 0.670 & Fitted \\
 & \texttt{lightextcoeff} & 0.5 & 0.5 & Fixed \\
 & \texttt{gmin} & 1.2 & 1.2 & Fixed \\
 & \texttt{b} & 0.031 & 0.08 & Fitted \\
 & \texttt{respcoeff} & 0.8 & 0.8 & Fixed \\
 & \texttt{intc} & 0.02 & 0.02 & Fixed \\
 & \texttt{knstore} & 0.1 & 0.1 & Fixed \\
 & \texttt{albedo\_leaf} & 0.18 & 0.18 & Fixed \\
 & \texttt{basetemp\_low} & 5 & 5 & Fixed \\
 & \texttt{fphusen} & 0.75 & 0.75 & Fixed \\
\addlinespace[2pt]
Rice & \texttt{laimax} & 7 & 7.59 & Fitted \\
 & \texttt{sla} & 0.0431 & 0.0700 & Fitted \\
 & \texttt{alphaa} & 1 & 0.402 & Fitted \\
 & \texttt{lightextcoeff} & 0.5 & 0.997 & Fitted \\
 & \texttt{gmin} & 1 & 0.3 & Fitted \\
 & \texttt{b} & 0.031 & 0.0211 & Fitted \\
 & \texttt{respcoeff} & 0.8 & 0.8 & Fixed \\
 & \texttt{intc} & 0.02 & 0.02 & Fixed \\
 & \texttt{knstore} & 0.1 & 0.1 & Fixed \\
 & \texttt{albedo\_leaf} & 0.18 & 0.3 & Fitted \\
 & \texttt{basetemp\_low} & 8 & 8 & Fixed \\
 & \texttt{fphusen} & 0.8 & 0.8 & Fixed \\
\addlinespace[2pt]
Soybean & \texttt{laimax} & 5 & 7.17 & Fitted \\
 & \texttt{sla} & 0.0326 & 0.0361 & Fitted \\
 & \texttt{alphaa} & 1 & 0.755 & Fitted \\
 & \texttt{lightextcoeff} & 0.5 & 1.1 & Fitted \\
 & \texttt{gmin} & 1.2 & 0.3 & Fitted \\
 & \texttt{b} & 0.031 & 0.0707 & Fitted \\
 & \texttt{respcoeff} & 0.8 & 1.5 & Fitted \\
 & \texttt{intc} & 0.02 & 0.02 & Fixed \\
 & \texttt{knstore} & 0.1 & 0.300 & Fitted \\
 & \texttt{albedo\_leaf} & 0.18 & 0.3 & Fitted \\
 & \texttt{basetemp\_low} & 7 & 7 & Fixed \\
 & \texttt{fphusen} & 0.7 & 0.760 & Fitted \\
\addlinespace[2pt]
\hline
\end{longtable}

\begin{longtable}{p{2.1cm}p{3.0cm}r@{\hspace{0.8cm}}rp{1.8cm}}
\caption{Final full-data soil parameterization used in the global simulations. Long values are shown to three significant digits. Simulations use the full-precision values.}\label{tab:final_soil_parameterization}\\
\hline
Parameter product & Parameter & Default & Final & Status \\
\hline
\endfirsthead
\multicolumn{5}{l}{\tablename\ \thetable\ continued}\\
\hline
Parameter product & Parameter & Default & Final & Status \\
\hline
\endhead
Winter wheat & \texttt{k\_soil10\_fast} & $1.10 \times 10^{-4}$ & $1.53 \times 10^{-4}$ & Fitted \\
 & \texttt{PRIESTLEY\_TAYLOR} & 1.32 & 0.8 & Fitted \\
 & \texttt{soildepth\_evap} & 300 & 175 & Fitted \\
 & \texttt{soil\_infil} & 2 & 1.01 & Fitted \\
\addlinespace[2pt]
Spring wheat & \texttt{k\_soil10\_fast} & $1.10 \times 10^{-4}$ & $1.00 \times 10^{-4}$ & Fitted \\
 & \texttt{PRIESTLEY\_TAYLOR} & 1.32 & 0.8 & Fitted \\
 & \texttt{soildepth\_evap} & 300 & 174 & Fitted \\
 & \texttt{soil\_infil} & 2 & 1 & Fitted \\
\addlinespace[2pt]
Maize & \texttt{k\_soil10\_fast} & $1.10 \times 10^{-4}$ & $3.00 \times 10^{-4}$ & Fitted \\
 & \texttt{PRIESTLEY\_TAYLOR} & 1.32 & 0.8 & Fitted \\
 & \texttt{soildepth\_evap} & 300 & 176 & Fitted \\
 & \texttt{soil\_infil} & 2 & 1 & Fitted \\
\addlinespace[2pt]
Rice & \texttt{k\_soil10\_fast} & $1.10 \times 10^{-4}$ & $2.14 \times 10^{-4}$ & Fitted \\
 & \texttt{PRIESTLEY\_TAYLOR} & 1.32 & 0.8 & Fitted \\
 & \texttt{soildepth\_evap} & 300 & 190 & Fitted \\
 & \texttt{soil\_infil} & 2 & 3.00 & Fitted \\
\addlinespace[2pt]
Soybean & \texttt{k\_soil10\_fast} & $1.10 \times 10^{-4}$ & $1.10 \times 10^{-4}$ & Fixed \\
 & \texttt{PRIESTLEY\_TAYLOR} & 1.32 & 1.32 & Fixed \\
 & \texttt{soildepth\_evap} & 300 & 300 & Fixed \\
 & \texttt{soil\_infil} & 2 & 2 & Fixed \\
\addlinespace[2pt]
\hline
\end{longtable}

\clearpage
\section{Global historical yield evaluation}

\begin{figure}[H]
\centering
\includegraphics[width=1\textwidth]{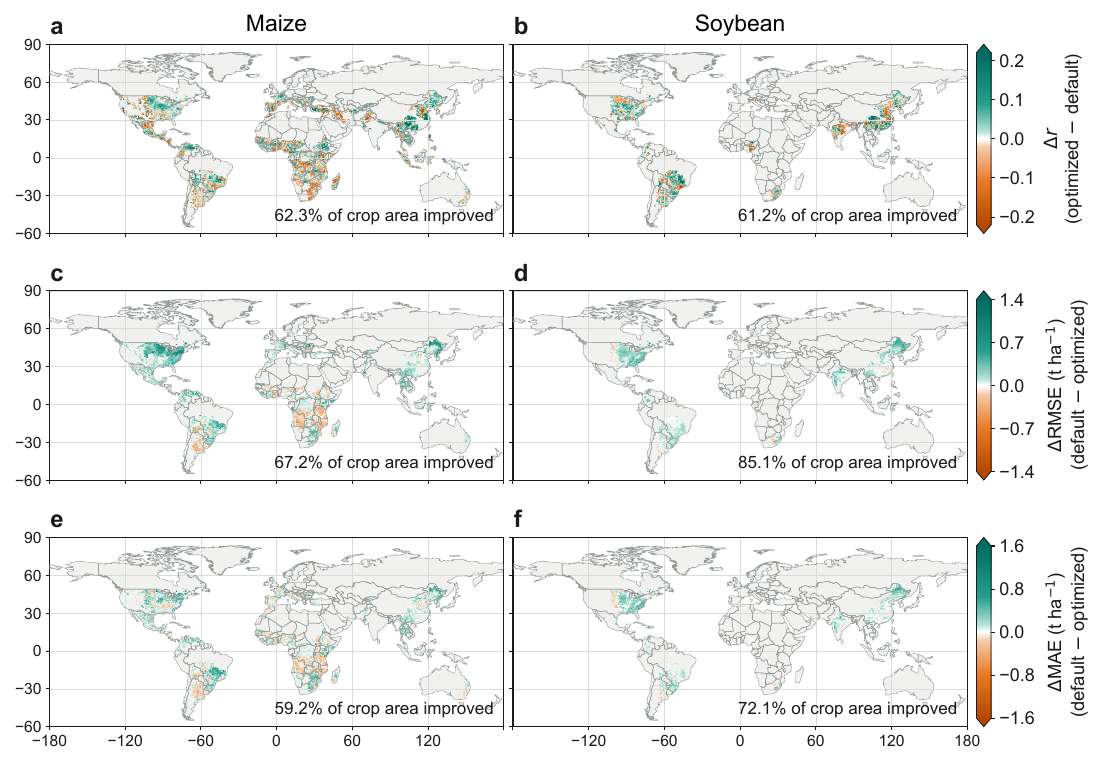}
\caption{\textbf{Site-trained parameters improve historical yield anomaly simulations and low-yield performance.}
\textbf{a, b}, Change in grid-cell temporal correlation between simulations and the global dataset of historical yields (GDHY) \cite{iizumi2020global_sup} for maize and soybean, calculated as $r_{\mathrm{optimized}}-r_{\mathrm{default}}$. 
\textbf{c, d}, Reduction in root-mean-square error (RMSE) for maize and soybean, calculated as $\mathrm{RMSE}_{\mathrm{default}}-\mathrm{RMSE}_{\mathrm{optimized}}$. 
\textbf{e, f}, Spatial changes in mean absolute error (MAE) in years belonging to the lowest 20\% of GDHY anomalies for maize and soybean after parameter optimization. Positive values indicate better performance in all panels. Percentages report the crop area with improved performance. Metrics use grid cells containing at least 10~ha of the crop area and at least five-year yield during 1981--2016.}
\label{fig:historical_yield_skill_maize_soybean}
\end{figure}

\begin{figure}[H]
\centering
\includegraphics[width=0.9\textwidth]{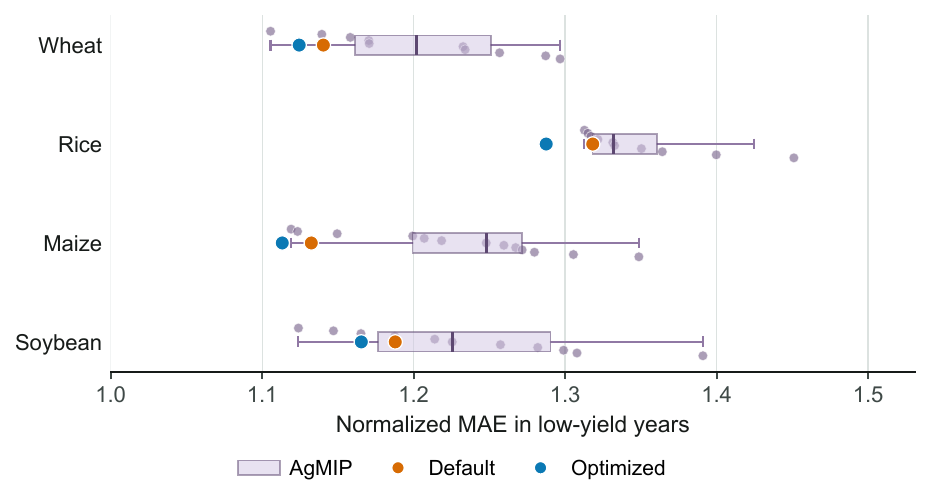}
\caption{\textbf{Observation-based parameter calibration improves historical low-yield simulations.}
Comparison of normalized mean absolute error (nMAE) for wheat, rice, maize and soybean in years belonging to the lowest 20\% of detrended GDHY yield anomalies during 1981--2016. Purple points show individual AgMIP crop models. Boxes show their median and inter-quartile range, with whiskers extending up to 1.5 times the interquartile range. Orange and blue points show the default and optimized simulations, respectively. Lower values indicate better performance.}
\label{fig:historical_low_yield_nmae_benchmark}
\end{figure}

\begin{table*}[ht]
\centering
\caption{Country-level evaluation of historical yield anomalies for the five largest producing countries of each crop. Countries are ranked by mean annual production for 1981--2016, calculated as grid-cell mean GDHY yield multiplied by crop area. Country metrics are crop area-weighted means of the grid-cell metrics. The $\Delta r$ is calculated as $r_{\mathrm{optimized}}-r_{\mathrm{default}}$, whereas $\Delta\mathrm{nRMSE}$ is calculated as $\mathrm{nRMSE}{\mathrm{optimized}}-\mathrm{nRMSE}{\mathrm{default}}$. Positive $\Delta r$ and negative $\Delta\mathrm{nRMSE}$ indicate improvement.}
\label{tab:top5_country_yield_skill}
\small
\setlength{\tabcolsep}{5pt}
\begin{tabular}{llrrrrrr}
\hline
Crop & Country & \multicolumn{3}{c}{Temporal correlation} & \multicolumn{3}{c}{nRMSE} \\
 & & Default & Optimized & $\Delta r$ & Default & Optimized & $\Delta$ \\
\hline
Wheat & United States & 0.353 & 0.389 & 0.037 & 1.692 & 0.976 & -0.716 \\
 & France & 0.054 & 0.130 & 0.076 & 1.337 & 0.982 & -0.355 \\
 & China & -0.061 & -0.048 & 0.013 & 1.214 & 1.014 & -0.200 \\
 & Canada & 0.321 & 0.388 & 0.067 & 1.126 & 0.932 & -0.194 \\
 & Germany & 0.322 & 0.436 & 0.114 & 1.228 & 0.912 & -0.316 \\
\addlinespace[2pt]
Rice & China & 0.094 & 0.160 & 0.066 & 1.090 & 1.009 & -0.081 \\
 & India & 0.050 & 0.080 & 0.030 & 1.119 & 1.037 & -0.082 \\
 & Thailand & 0.260 & 0.246 & -0.014 & 1.407 & 0.963 & -0.444 \\
 & Bangladesh & 0.064 & 0.063 & -0.001 & 1.083 & 1.032 & -0.051 \\
 & Indonesia & -0.054 & 0.063 & 0.117 & 1.231 & 1.021 & -0.209 \\
\addlinespace[2pt]
Maize & United States & 0.198 & 0.237 & 0.040 & 2.639 & 2.209 & -0.430 \\
 & China & 0.176 & 0.212 & 0.036 & 0.994 & 0.858 & -0.136 \\
 & Brazil & 0.283 & 0.294 & 0.011 & 1.116 & 0.967 & -0.149 \\
 & Argentina & 0.541 & 0.524 & -0.018 & 1.114 & 1.308 & 0.194 \\
 & France & 0.515 & 0.543 & 0.028 & 1.162 & 1.050 & -0.112 \\
\addlinespace[2pt]
Soybean & United States & 0.296 & 0.319 & 0.023 & 2.173 & 1.861 & -0.312 \\
 & Brazil & 0.124 & 0.175 & 0.051 & 1.162 & 1.075 & -0.087 \\
 & Argentina & 0.519 & 0.543 & 0.024 & 0.940 & 0.903 & -0.036 \\
 & India & 0.374 & 0.361 & -0.013 & 1.111 & 0.958 & -0.153 \\
 & China & 0.102 & 0.091 & -0.011 & 1.916 & 1.604 & -0.312 \\
\hline
\end{tabular}
\end{table*}

\clearpage
\begin{figure}[H]
\centering
\makebox[\textwidth][c]{
\includegraphics[width=1.2\textwidth]{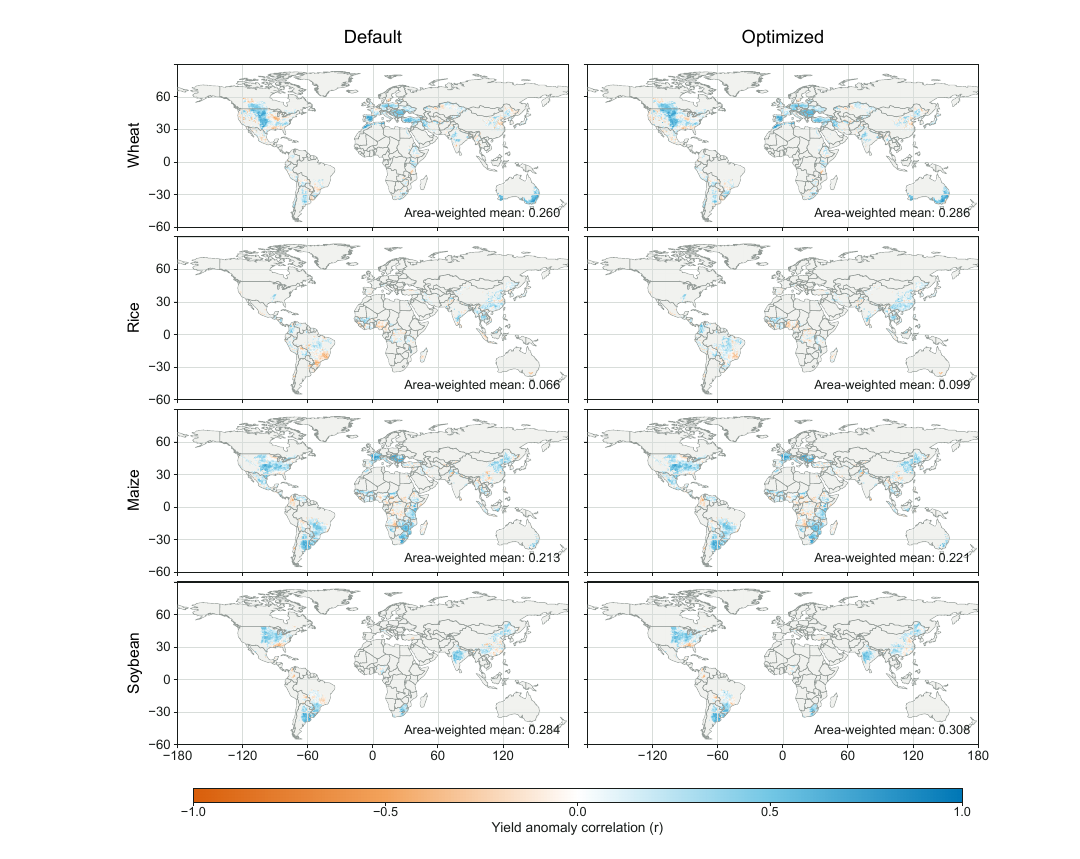}
}
\caption{\textbf{Temporal correlation of historical yield anomalies.}
At each grid cell, simulated yield time series and the yield time series from the global dataset of historical yields (GDHY) \cite{iizumi2020global_sup} are detrended independently over 1981--2016. Temporal correlations are then calculated between the resulting anomalies. Rows show wheat, rice, maize and soybean. Columns show the default and optimized simulations. Maps show grid cells with at least 10~ha of cultivated area for the crop and at least five years of yield data.}
\label{fig:absolute_yield_correlation}
\end{figure}

\begin{figure}[H]
\centering
\makebox[\textwidth][c]{
\includegraphics[width=1.2\textwidth]{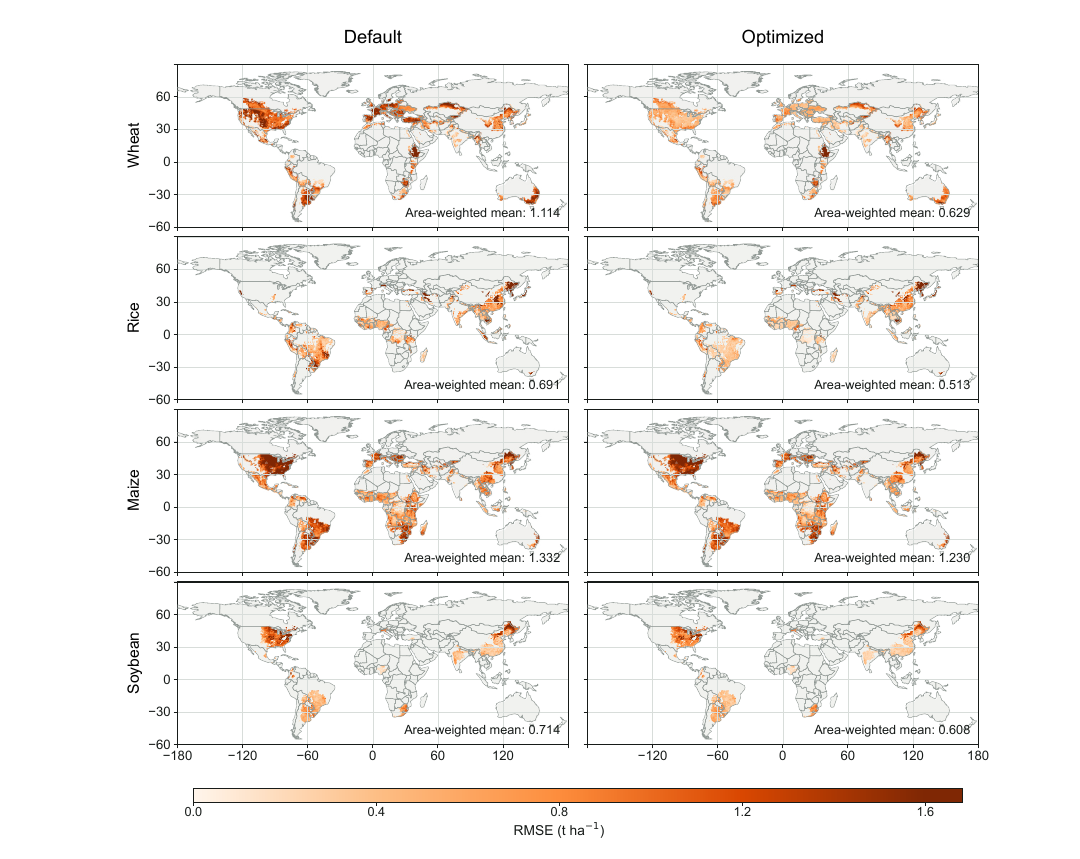}
}
\caption{\textbf{Absolute root-mean-square error (RMSE) of historical yield anomalies.}
Rows show wheat, rice, maize and soybean. Columns show the default and optimized simulations. The RMSE is calculated between independently detrended simulated yield anomalies and yield anomalies from the global dataset of historical yields (GDHY) for 1981--2016 \cite{iizumi2020global_sup}, then normalized by the temporal standard deviation of the GDHY yield-anomaly series. Lower values indicate better agreement. Maps retain grid cells containing at least 10~ha cultivated area of the crop and at least five-year yield data.}
\label{fig:absolute_yield_nrmse}
\end{figure}

\begin{figure}[H]
\centering
\includegraphics[width=0.92\textwidth]{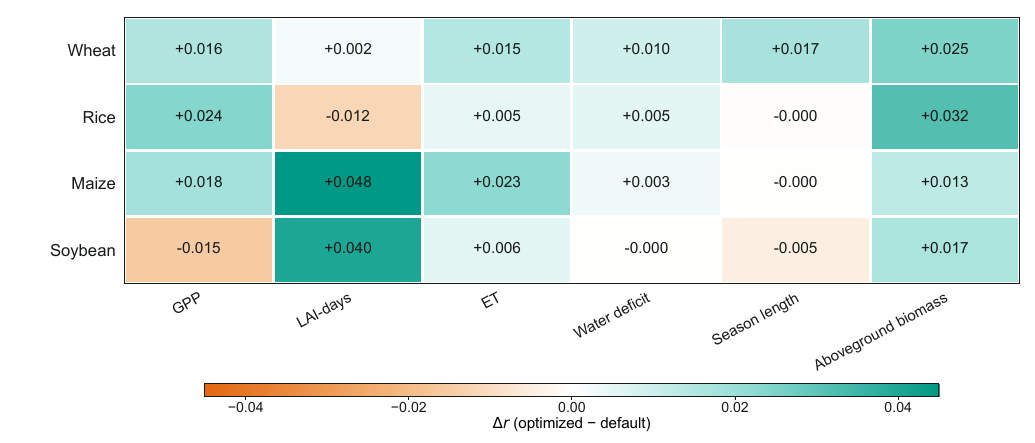}
\caption{\textbf{Changes in seasonal process--yield correlations.}
Values show changes in the crop area-weighted mean correlation after parameter optimization. Correlations are calculated over 1981--2016 between yield anomalies from the global dataset of historical yields (GDHY) \cite{iizumi2020global_sup} and anomalies in six simulated seasonal variables: gross primary productivity (GPP), LAI-days (the sum of daily leaf area index over the growing season), evapotranspiration (ET), water deficit, growing-season length and aboveground biomass. Positive values indicate higher correlations with the optimized parameters.}
\label{fig:process_yield_correlations}
\end{figure}

\begin{center}
\begin{minipage}{\textwidth}
\centering
\small
\setlength{\tabcolsep}{3.5pt}
\captionof{table}{\textbf{Differences in end-of-century global production responses between optimized and default parameterizations across climate models.}
The differences between optimized and default simulations in the 2080--2099 mean production response, expressed in percentage points. Each response is calculated from grid-cell yield changes relative to the corresponding 1995--2014 mean and aggregated using crop-area weights. Positive and negative values indicate more positive and more negative responses, respectively, with optimized parameters. The final column reports the number of climate models with the indicated sign.}
\label{tab:climate_model_parameterization_effects}
\begin{tabular}{llrrrrrl}
\hline
& & \multicolumn{5}{c}{Response difference (percentage points)} & \\
Pathway & Crop & GFDL & IPSL & MPI & MRI & UKESM & Sign agreement \\
\hline
SSP1--2.6 & Wheat   & $-0.46$ & $+1.84$ & $+1.09$ & $+0.75$ & $+2.85$ & 4/5 positive \\
          & Rice    & $-4.57$ & $-9.20$ & $-6.23$ & $-2.66$ & $-2.73$ & 5/5 negative \\
          & Maize   & $+0.42$ & $+0.84$ & $+0.38$ & $+0.88$ & $+1.92$ & 5/5 positive \\
          & Soybean & $-0.51$ & $+0.08$ & $+0.04$ & $+2.34$ & $+3.47$ & 4/5 positive \\
\addlinespace[2pt]
SSP5--8.5 & Wheat   & $+3.88$ & $+6.94$ & $+5.10$ & $+4.39$ & $+6.05$ & 5/5 positive \\
          & Rice    & $-48.04$ & $-32.83$ & $-39.66$ & $-32.97$ & $-32.65$ & 5/5 negative \\
          & Maize   & $+1.76$ & $+3.25$ & $+2.94$ & $+1.62$ & $+4.52$ & 5/5 positive \\
          & Soybean & $-1.16$ & $+0.38$ & $-0.001$ & $-3.97$ & $+1.75$ & 3/5 negative \\
\hline
\end{tabular}
\end{minipage}
\end{center}

\begin{figure}[H]
\centering
\makebox[\textwidth][c]{
\includegraphics[width=1.2\textwidth]{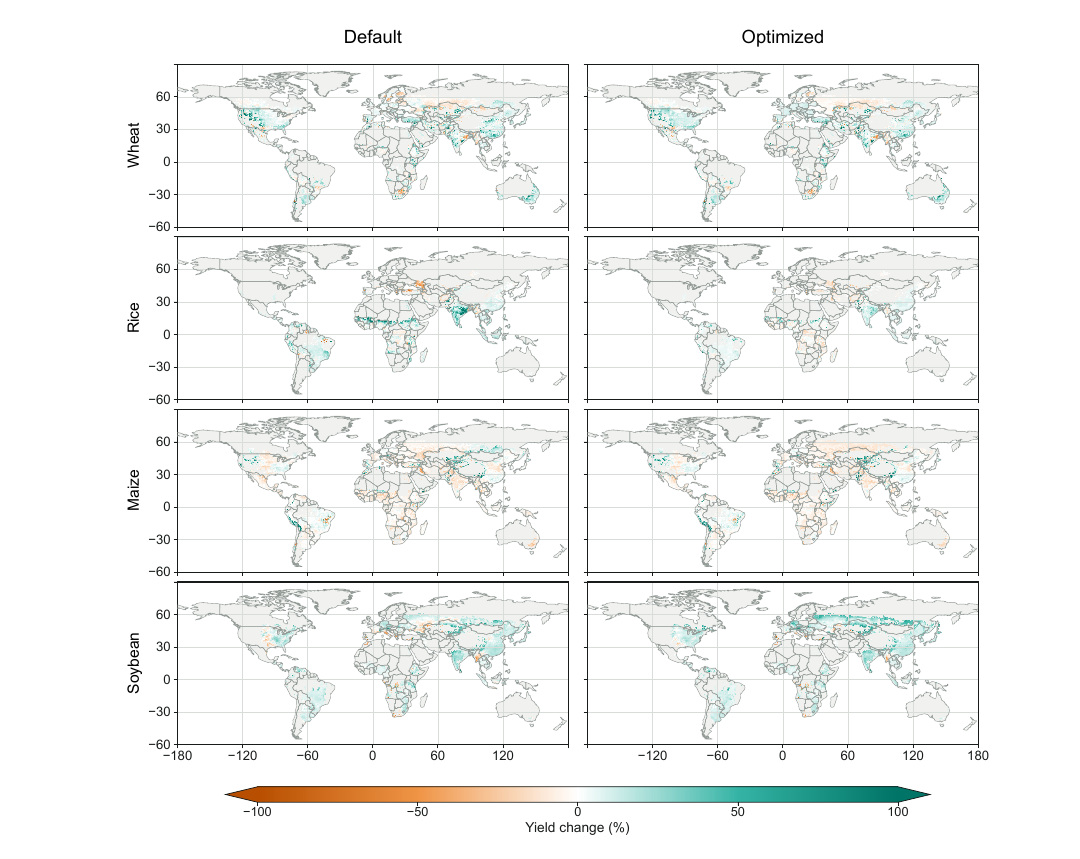}
}
\caption{\textbf{Spatial distribution of end-of-century simulated yield changes under SSP1--2.6.}
Rows show wheat, rice, maize and soybean. Columns show the default and optimized parameterizations. For each parameterization and climate model, grid-cell yield is averaged over 2080--2099 and expressed relative to its own 1995--2014 mean. Maps show the median response across five climate models, i.e., GFDL-ESM4, IPSL-CM6A-LR, MPI-ESM1-2-HR, MRI-ESM2-0 and UKESM1-0-LL. Rainfed and irrigated yields are combined using crop area.}
\label{fig:future_yield_maps_end_of_century_ssp126}
\end{figure}

\begin{figure}[H]
\centering
\makebox[\textwidth][c]{
\includegraphics[width=1.2\textwidth]{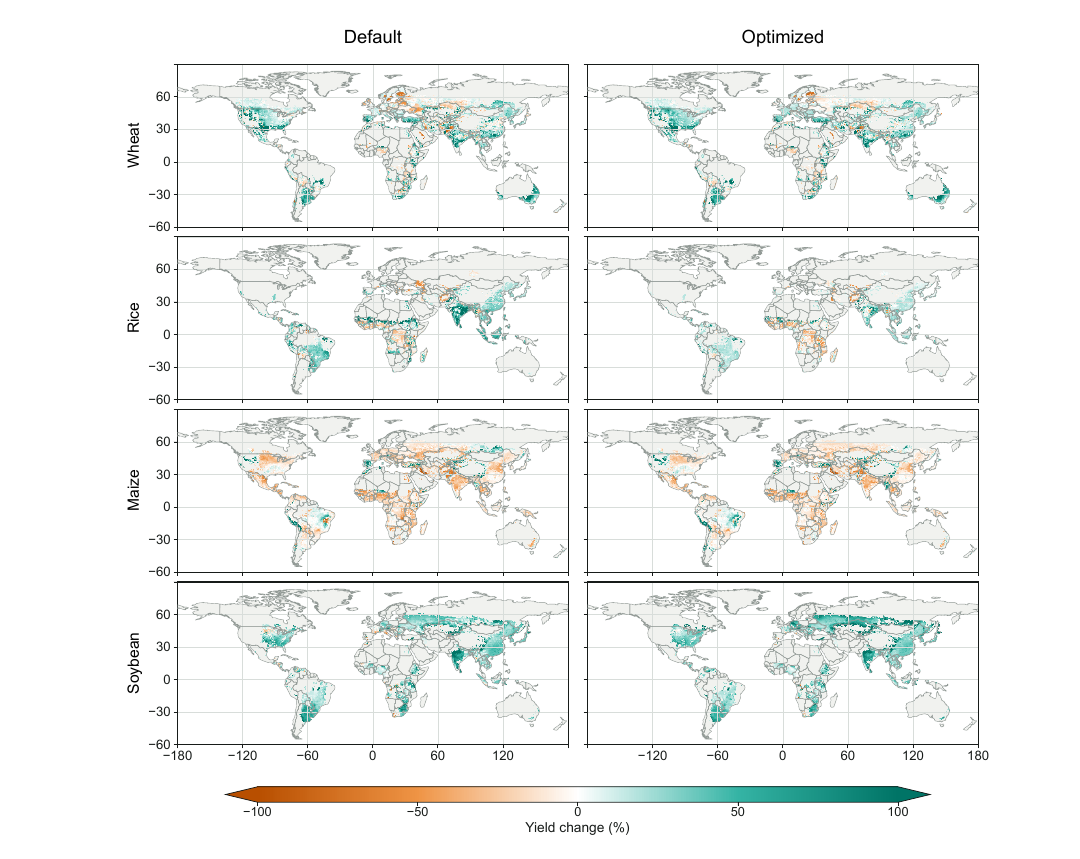}
}
\caption{\textbf{Spatial distribution of end-of-century simulated yield changes under SSP5--8.5.}
Rows show wheat, rice, maize and soybean. Columns show the default and optimized parameterizations. For each parameterization and climate model, grid-cell yield is averaged over 2080--2099 and expressed relative to its own 1995--2014 mean. Maps show the median response across five climate models, i.e., GFDL-ESM4, IPSL-CM6A-LR, MPI-ESM1-2-HR, MRI-ESM2-0 and UKESM1-0-LL. Rainfed and irrigated yields are combined using crop area.}
\label{fig:future_yield_maps_end_of_century_ssp585}
\end{figure}

\clearpage
\section{Parameter-ensemble diagnostics}\label{sec:parameter_ensemble_supp}

\begin{center}
\begin{minipage}{\textwidth}
\centering
\small
\captionof{table}{\textbf{Historical performance and future response across the four crop parameter ensembles.}
Historical metrics are calculated for the default, optimized and all 20 bootstrap parameterizations. Bootstrap rows report the median and, in parentheses, the 10th--90th percentile interval. End-of-century responses are calculated as the 2080--2099 mean for each parameterization--climate-model combination relative to its own 1995--2014 simulations. Default and optimized rows report the median across five climate models. For each bootstrap member, we first calculate the five-model median. Bootstrap rows report the median and 10th--90th percentile range across the 20 members.}
\label{tab:four_crop_parameter_ensemble_skill}
\setlength{\tabcolsep}{3pt}
\begin{tabular}{llrrr}
\multicolumn{5}{l}{\textbf{a, Historical performance}} \\[2pt]
\hline
Crop & Parameterization & Correlation & nRMSE & Low-yield MAE \\
\hline
Wheat & Default & 0.260 & 1.113 & 3.006 \\
 & Optimized & 0.286 & 0.628 & 1.699 \\
 & Bootstrap, $n=20$ & 0.284 (0.281--0.287) & 0.627 (0.622--0.635) & 1.697 (1.685--1.723) \\
\noalign{\vskip 2pt}
Rice & Default & 0.066 & 0.691 & 1.789 \\
 & Optimized & 0.099 & 0.513 & 1.414 \\
 & Bootstrap, $n=20$ & 0.096 (0.085--0.104) & 0.537 (0.502--0.599) & 1.458 (1.401--1.531) \\
\noalign{\vskip 2pt}
Maize & Default & 0.213 & 1.332 & 2.320 \\
 & Optimized & 0.227 & 1.195 & 2.242 \\
 & Bootstrap, $n=20$ & 0.225 (0.222--0.229) & 1.158 (1.062--1.266) & 2.142 (1.964--2.327) \\
\noalign{\vskip 2pt}
Soybean & Default & 0.284 & 0.713 & 2.872 \\
 & Optimized & 0.308 & 0.608 & 2.431 \\
 & Bootstrap, $n=20$ & 0.300 (0.295--0.308) & 0.625 (0.515--0.765) & 2.491 (2.041--3.054) \\
\hline
\end{tabular}

\vspace{6pt}

\begin{tabular}{llrr}
\multicolumn{4}{l}{\textbf{b, End-of-century production response (\%)}} \\[2pt]
\hline
Crop & Parameterization & SSP1--2.6 & SSP5--8.5 \\
\hline
Wheat & Default & 4.95 & 12.81 \\
 & Optimized & 5.70 & 17.91 \\
 & Bootstrap, $n=20$ & 5.60 (5.17--5.76) & 17.14 (15.09--18.78) \\
\noalign{\vskip 2pt}
Rice & Default & 12.16 & 55.08 \\
 & Optimized & 6.08 & 22.11 \\
 & Bootstrap, $n=20$ & 7.47 (5.68--10.29) & 29.23 (21.83--49.35) \\
\noalign{\vskip 2pt}
Maize & Default & $-2.82$ & $-15.48$ \\
 & Optimized & $-1.98$ & $-12.54$ \\
 & Bootstrap, $n=20$ & $-1.82$ ($-2.13$--$-1.56$) & $-15.22$ ($-18.47$--$-13.49$) \\
\noalign{\vskip 2pt}
Soybean & Default & 10.93 & 37.44 \\
 & Optimized & 11.00 & 37.43 \\
 & Bootstrap, $n=20$ & 10.76 (9.99--12.93) & 37.21 (36.15--42.30) \\
\hline
\end{tabular}
\end{minipage}
\end{center}

\begin{center}
\begin{minipage}{\textwidth}
\centering
\small
\captionof{table}{\textbf{Contributions of parameterization and climate models to variation in projected production responses.}
All crops and pathways use the same 20 bootstrap parameterizations. The global production response is averaged over 2080--2099. A two-factor analysis of variance (ANOVA) partitions the resulting variation into parameterization, climate-model and interaction components.}
\label{tab:parameter_gcm_variance}
\resizebox{\textwidth}{!}{
\begin{tabular}{llrrrr}
\hline
Pathway & Crop & Parameter members & Parameter set (\%) & Climate model (\%) & Interaction (\%) \\
\hline
SSP1--2.6 & Wheat & 20 & 0.8 & 99.0 & 0.3 \\
 & Rice & 20 & 28.1 & 63.0 & 8.9 \\
 & Maize & 20 & 1.2 & 97.4 & 1.4 \\
 & Soybean & 20 & 8.0 & 90.2 & 1.8 \\
\addlinespace[2pt]
SSP5--8.5 & Wheat & 20 & 0.9 & 99.0 & 0.1 \\
 & Rice & 20 & 63.3 & 31.6 & 5.1 \\
 & Maize & 20 & 5.6 & 93.0 & 1.4 \\
 & Soybean & 20 & 26.1 & 69.7 & 4.2 \\
\hline
\end{tabular}
}
\end{minipage}
\end{center}

\begin{figure}[p]
\centering
\makebox[\textwidth][c]
{\includegraphics[width=1.0\textwidth]{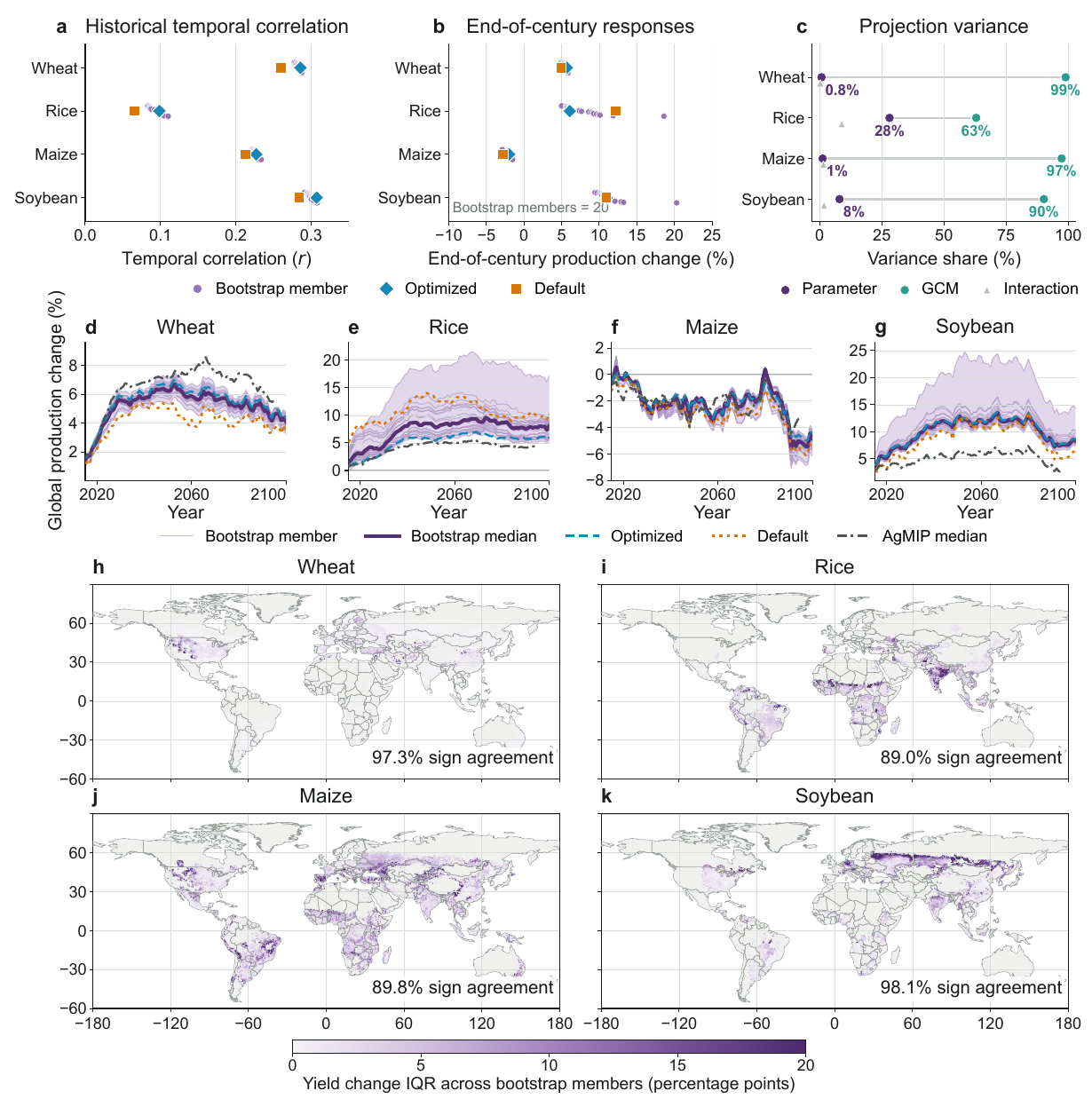}
}
\caption{\textbf{Parameterizations with similar historical performance produce different future projections.}
Like Fig. 4 but for SSP1--2.6 instead of SSP5--8.5.
\textbf{a}, Historical temporal correlation ($r$) against the global dataset of historical yields (GDHY) for 1981--2016 \cite{iizumi2020global_sup}.
\textbf{b}, End-of-century (2080--2099) global production changes summarized by the five-model median. In \textbf{a, b}, purple points show individual bootstrap members. Blue diamonds and orange squares show the optimized parameterization calibrated using all sites and the default parameterization, respectively.
\textbf{c}, Variance shares in end-of-century global production changes attributable to parameters, climate models and their interaction.
\textbf{d--g}, Global production trajectories. Thin purple lines show the five-model medians of individual members. The thick purple line shows the ensemble median, and shading spans the range of the smoothed bootstrap-member curves. Blue and orange dash lines show the optimized and default parameterizations. The grey line shows the AgMIP ensemble median. Curves use centered 11-year moving averages.
\textbf{h--k}, Interquartile range ($Q_{75}-Q_{25}$) of grid-cell yield changes for 2080--2099 relative to 1995--2014 across bootstrap members, calculated from their five-model median responses and expressed in percentage points. Values above 20 percentage points use the upper colour limit. Annotations report the percentage of crop area where at least 17 of the 20 bootstrap members agree on the sign of yield change.}
\label{fig:parameter_uncertainty_ssp126}
\end{figure}

\clearpage
\putbib[sn-bibliography-sup]
\end{bibunit}